\documentclass[lettersize,journal]{IEEEtran}
\usepackage{cite}
\usepackage{amsmath,amssymb,amsfonts,amsthm}
\usepackage{comment}

\usepackage{algorithmic}
\usepackage{algorithm}
\usepackage{enumerate}  
\DeclareMathOperator{\Tr}{Tr}
\DeclareMathOperator{\diag}{diag}

\usepackage{url}
\usepackage{cite}
\usepackage[english]{babel}
\newtheorem{lemma}{Lemma}
\newtheorem{remark}{Remark}

\usepackage{subcaption}
\usepackage{graphicx}

\usepackage{textcomp}
\usepackage{optidef}
\usepackage{xcolor}

\def\BibTeX{{\rm B\kern-.05em{\sc i\kern-.025em b}\kern-.08em
    T\kern-.1667em\lower.7ex\hbox{E}\kern-.125emX}}
\usepackage{balance}

\begin{document}

\title{Near-Field Integrated Sensing, Computing and Semantic Communication in Digital Twin-Assisted Vehicular Networks}

\author{Yinchao Yang, Yahao Ding, Jiaxiang Wang, Zhaohui Yang, Chen Zhu, Zhaoyang Zhang, Dusit Niyato, \IEEEmembership{Fellow, IEEE}, and Mohammad Shikh-Bahaei,
\IEEEmembership{Senior Member, IEEE}
\thanks{Copyright (c) 20xx IEEE. Personal use of this material is permitted. However, permission to use this material for any other purposes must be obtained from the IEEE by sending a request to pubs-permissions@ieee.org.}
\thanks{Yinchao Yang, Yahao Ding, Jiaxiang Wang, and Mohammad Shikh-Bahaei are with the Department of Engineering, King's College London, London, UK (emails: yinchao.yang@kcl.ac.uk; jiaxiang.wang@kcl.ac.uk; yahao.ding@kcl.ac.uk; m.sbahaei@kcl.ac.uk).}
\thanks{Zhaohui Yang, Chen Zhu, and Zhaoyang Zhang are with the College of Information Science and Electronic Engineering, Zhejiang University, Hangzhou, Zhejiang 310027, China, and Zhejiang Provincial Key Lab of Information Processing, Communication and Networking (IPCAN), Hangzhou, Zhejiang, 310007, China (email: yang\_zhaohui@zju.edu; zhuc@zju.edu.cn; ning\_ming@zju.edu.cn).}
\thanks{Dusit Niyato is with the College of Computing and Data Science, Nanyang Technological University, Singapore 639798, Singapore (email: dniyato@ntu.edu.sg).}
}

\markboth{Journal of \LaTeX\ Class Files,~Vol.~18, No.~9, September~2020}%
{How to Use the IEEEtran \LaTeX \ Templates}

\maketitle

\begin{abstract}
    The rapid development of digital twin (DT) technology has introduced transformative potential for vehicular networks, enabling real-time, high-fidelity virtual representations that can enhance safety, efficiency, and automation. However, realizing seamless DT synchronization in dynamic vehicular environments presents significant challenges, including the need for high data rates to support massive data transmission, precision sensing for accurate environmental mapping, and efficient resource management under strict latency and computational constraints. Addressing these challenges, this paper proposes an integrated sensing, computing, and semantic communication (ISCSC) framework specifically tailored for DT-assisted vehicular networks operating in the near-field (NF) regime. Leveraging a multi-user multiple-input multiple-output (MU-MIMO) configuration, each roadside unit (RSU) employs semantic communication to efficiently serve a subset of vehicles within the NF region while simultaneously utilizing millimeter-wave (mmWave) radar to detect all vehicles within its coverage area. We employ particle filtering at RSUs to achieve superior vehicle tracking accuracy. To optimize the system performance, we formulate a joint optimization problem balancing semantic communication rate and sensing accuracy with limited computational constraints. We developed a hybrid heuristic algorithm for vehicle-to-RSU assignments, followed by an alternating optimization approach for finding the optimal semantic extraction ratio and beamforming matrices. We extensively evaluate system performance through simulations, assessing the Crámer-Rao bound (CRB) for angle and distance estimation, semantic transmission rates, and computational resource utilization. Numerical results demonstrate that, under constrained resource conditions, the proposed ISCSC framework achieves a 20\% improvement in transmission rate while maintaining the sensing accuracy of existing ISAC schemes.
\end{abstract}


\section{Introduction}

\begin{table*}[!htbp]
    \centering
    \caption{Comparison of Existing Studies and the Proposed ISCSC Framework}
    \label{tab:comparison}
    \renewcommand{\arraystretch}{1.5}
    \begin{tabular}{|l|c|c|c|c|c|l|}
    \hline
    \textbf{Reference} & \textbf{ISAC} & \textbf{SC} & \textbf{NF Effect} & \textbf{DT} & \textbf{Computing}  & \textbf{Design Focus}  \\
    \hline\hline
    \cite{yang2024dynamic} & $\checkmark$ & $\times$ & $\times$ & $\times$ & $\times$ & Dynamic power allocation for ISAC-enabled vehicular networks\\
    \hline
    \cite{meng2023vehicular} & $\checkmark$ & $\times$ & $\times$ & $\times$ & $\times$& Beam tracking and trajectory-aware ISAC beam design \\
    \hline
    \cite{lin2024near} & $\checkmark$ & $\times$ & $\checkmark$ & $\times$ & $\times$& Near-field beamforming design for ISAC systems \\
    \hline
     \cite{zhao2024iov} & $\checkmark$ & $\times$ & $\times$ & $\times$ & $\checkmark$ &  ISAC-enabled IoV with mobile edge computing (MEC)\\
    \hline
    \cite{sha2024integrated} & $\checkmark$ & $\times$ & $\times$ & $\times$ & $\checkmark$ & Targeted dissemination strategies for ISAC and MEC-enabled IoV \\
    \hline
    \cite{su2023semantic} & $\times$ & $\checkmark$ & $\times$ & $\times$ & $\times$& Dynamic resource allocation for semantic communication \\
    \hline
    \cite{xia2023xurllc} & $\times$ & $\checkmark$ & $\times$ & $\times$ & $\times$ & Semantic-based URLLC optimization for vehicular communication \\
    \hline
    \cite{yang2023task} & $\times$ & $\checkmark$ & $\times$ & $\times$ & $\times$ & Energy-efficient semantic-aware cooperative transmission  \\
    \hline
    \cite{qin2024computing} & $\times$ & $\checkmark$ & $\times$ & $\times$ & $\checkmark$ &  Discussion of cloud computing for semantic communication\\
    \hline
    \cite{zhao2025compression} & $\times$ & $\checkmark$ & $\times$ & $\times$ & $\checkmark$ &  Resource allocation for probabilistic semantic communication with RSMA  \\
    \hline
    \cite{cui2023physical} & $\checkmark$ & $\times$ & $\times$ & $\checkmark$ & $\times$ & ISAC-enhanced DT modeling at the physical layer \\
    \hline
    \cite{gong2022resource} & $\checkmark$ & $\times$ & $\times$ & $\checkmark$ & $\times$ & Resource allocation for ISAC in DT-enabled IoV \\
    \hline
    \cite{zhang2024denoising} & $\checkmark$ & $\times$ & $\times$ & $\checkmark$ & $\times$ & ISAC-MIMO channel modeling with DT using diffusion probabilistic models \\
    \hline
    \cite{thomas2023causal} & $\times$ & $\checkmark$ & $\times$ & $\checkmark$ & $\times$ & End-to-end transmitter and receiver design \\
    \hline
    \cite{okegbile2025fles} & $\times$ & $\checkmark$ &$\times$ & $\checkmark$ & $\times$ & Integrating federated learning with SC \\
    \hline
   
    \textbf{This Work} & $\checkmark$ & $\checkmark$ & $\checkmark$ & $\checkmark$ & $\checkmark$ & Joint framework for DT-assisted vehicular networks with NF modeling\\
    \hline
    \end{tabular}
\end{table*}

\IEEEPARstart{D}{igital} twin (DT) technology is expected to play a key role in the development of future vehicular networks by enabling real-time monitoring, predictive analytics, and intelligent resource management \cite{tao2018digital,liu2024digital}. By creating accurate virtual models of vehicles and infrastructure, DTs help vehicular networks respond more effectively to changing conditions, improve system performance and safety, and enhance resource utilization. When combined with sixth-generation (6G) wireless networks, DTs can take advantage of high-frequency bands such as millimeter-wave (mmWave) and terahertz (THz), which offer higher sensing precision. However, using these high frequencies and large antenna arrays introduces strong near-field (NF) effects, where the radio waves no longer travel as planar waves but instead as spherical waves \cite{wang2023near}. These effects become significant within the Rayleigh distance, which is given by $R = \tfrac{2D^{2} f_{c}}{c}$, and it grows quadratically with the array aperture $D$ and linearly with the carrier frequency $f_{c}$. Hence, in typical 6G scenarios, the near-field region can extend up to several hundred meters \cite{an2024near}. For example, when $f_c = 50~\text{GHz}$ and $D = 1~\text{m}$, the Rayleigh distance reaches approximately $300~\text{m}$, indicating that most vehicles naturally fall within the RSU's NF zone to avoid high path loss and attenuation \cite{navdeti2022roadside}. As a result, accurately modeling and accounting for NF behavior is essential for fully realizing the benefits of DTs in high-frequency vehicular networks. Despite the advantages of operating in high-frequency bands, several key challenges remain for DT-enabled vehicular networks. First, even with the abundant spectrum, meeting the stringent and conflicting requirements of both communication and sensing functions remains difficult. Reliable communication requires high data rates and ultra-low latency, whereas sensing calls for high resolution and precise accuracy, both functions often compete for the same system resources \cite{tang2024joint}. Second, vehicular network scenarios demand intelligent and efficient data exchange mechanisms to support the massive and heterogeneous information flow between vehicles, infrastructure, and the DT platform \cite{gimenez2024semantic}. Conventional communication methods often involve the transmission of redundant data, leading to inefficient use of bandwidth and increased latency.

To address the competing demands of sensing and communication functions, integrated sensing and communication (ISAC) has emerged as a vital paradigm. ISAC aims to unify both functions, improving efficiency and capabilities by allowing both to operate seamlessly within the same system \cite{zhou2024near}. For instance, a base station (BS) can utilize communication signals for object sensing while leveraging sensing results to refine channel state information (CSI), enhancing both communication reliability and sensing accuracy \cite{11208653}. ISAC has attracted significant research attention in vehicular networks \cite{yang2024dynamic,meng2023vehicular,lin2024near,zhao2024iov,sha2024integrated}. In \cite{yang2024dynamic}, the authors proposed a dynamic power allocation strategy for ISAC-enabled vehicular networks, optimizing power distribution to maintain robust communication and accurate sensing under mobility constraints. The authors in \cite{meng2023vehicular} introduced a roadway-aware beam-tracking approach that incorporates roadway geometry to enhance beam alignment and connectivity for vehicles on complex trajectories. 
Other works have explored NF beamforming optimization for ISAC systems~\cite{lin2024near}, as well as the extension to integrated sensing, communication, and computation (ISCC) frameworks~\cite{zhao2024iov,sha2024integrated}, aiming to improve the feasibility of real-world deployments.

To fulfill the requirement for intelligent data exchange, semantic communication (SC) has been introduced into DT-enabled vehicular networks \cite{ding2023distributed}. Unlike conventional communication, which is fundamentally constrained by Shannon’s capacity theorem, SC optimizes resource utilization by prioritizing the semantic relevance of transmitted data, thereby significantly enhancing efficiency and reducing redundant overhead \cite{li2024secure, wang2025generative}. 
SC achieves this by exploiting shared knowledge bases (KBs) at both the transmitter and the receiver, which encapsulate common knowledge, ground truth, and prior interactions. Leveraging these KBs, the receiver can reconstruct or infer the intended meaning of a message, even when only partial, abstract, or compressed semantic representations are transmitted. Extensive research has been conducted in this area \cite{su2023semantic,xia2023xurllc,yang2023task,qin2024computing,zhao2025compression}. In \cite{su2023semantic}, the authors proposed a semantic-aware resource allocation framework for device-to-device (D2D) vehicular networks, optimizing spectrum efficiency while preserving essential semantic information. Similarly, \cite{xia2023xurllc} presented an xURLLC-aware service provisioning framework while minimizing unnecessary data transmission. Furthermore, \cite{yang2023task} developed a task-driven, semantic-aware cooperative transmission strategy for vehicular networks, reducing energy consumption while ensuring contextually relevant and reliable information exchange. 
Several studies, such as those in \cite{qin2024computing,zhao2025compression}, have integrated computing models and computing resources into their overall design framework.

Nevertheless, the integration of SC with ISAC for DT-assisted vehicular networks, particularly under the NF effect, remains largely unexplored. Within vehicular networks, DTs are typically instantiated and maintained by roadside units (RSUs). Each RSU is tasked with collecting dynamic vehicular information, such as position, velocity, and trajectory, from nearby vehicles within its service range. Upon receiving the data, the RSU performs a series of signal processing tasks, including filtering, feature extraction, multi-source data fusion, and building the DT models \cite{wang2022mobility}. With this digital environment in place, RSUs can perform proactive, context-aware decision-making to support a range of intelligent transportation functions. These include traffic flow optimization, safety-critical event prediction, and cooperative vehicle manoeuvres such as collision avoidance, lane-change coordination, and adaptive speed control \cite{dai2022adaptive}. The resulting decisions are then communicated back to vehicles and, when necessary, shared among adjacent RSUs to ensure consistent and coordinated vehicular control across the network. To enable timely data exchange, SC allows the RSUs to prioritize critical information and eliminate redundant data before transmission \cite{wang2025deep}. Consequently, environmental sensing, semantic information extraction and sharing, and the computational capabilities required for building DT models must all be carefully considered in the overall system design.

\begin{figure}[!t]
    \centering
    \includegraphics[width=0.6\linewidth]{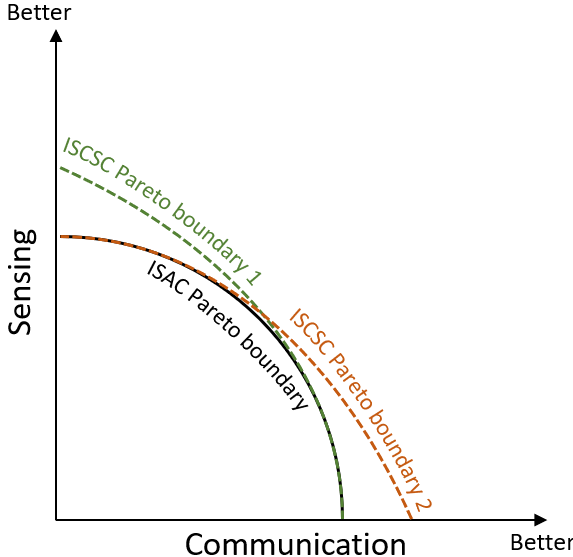}
    \caption{The Pareto boundaries for the ISAC and the proposed ISCSC systems. The ISCSC Pareto boundary 1 and 2 correspond to more improvement on sensing than communication, and more improvement on communication than sensing, respectively.}
    \label{fig:pareto comp}
\end{figure}

While the integration of ISAC with DT \cite{cui2023physical, gong2022resource, zhang2024denoising}, and the integration of SC with DT \cite{thomas2023causal, okegbile2025fles}, have been individually studied, their combined integration remains an open research challenge. The main difficulty is that sensing, semantic communication, and DT modeling all compete for the same limited power and processing ability. As a result, a unified approach is essential to manage the difficult trade-offs between ensuring high sensing accuracy, achieving reliable communication, and handling the computational demands and time delays of creating and maintaining an accurate DT. Motivated by these gaps, this paper introduces a novel integrated sensing, computing, and semantic communication (ISCSC) framework for DT-enabled vehicular networks. As shown in Fig.~\ref{fig:pareto comp}, the proposed framework effectively expands the Pareto boundary, offering better sensing performance without compromising communication (Pareto boundary~1), or better communication performance without sacrificing sensing (Pareto boundary~2). This capability is essential for future ISAC systems that must function in dynamic environments where sensing accuracy and communication reliability are equally critical. Conventional ISAC designs are often constrained by limited resources, while the proposed framework enhances the overall system capability within the same limitations, ensuring more efficient and adaptive operation.

To clearly highlight the distinctions between our proposed framework and existing state-of-the-art works, we summarize key studies along with their design focuses in Table~\ref{tab:comparison}. This comparison underscores the novelty of our work in integrating sensing, semantic communication, near-field effects, and digital twin technologies within vehicular networks. The main contributions of this work are as follows:
\begin{enumerate}[\roman{enumi})]
    \item We propose the first integrated ISCSC framework explicitly accounting for NF effects in DT-enabled vehicular networks with multiple RSUs and multiple vehicles. Each RSU and vehicle is equipped with a multi-antenna array, enabling multi-user multiple-input multiple-output (MU-MIMO) transmission and reception, in line with practical deployments in vehicular networks.

    \item We introduce a novel computational model that establishes a direct relationship between the computational resource required for building a DT model at the RSU and sensing accuracy, quantified by the root Cramér-Rao bound (RCRB). This model provides an estimation of the additional computational workload required to compensate for sensing errors, enabling more efficient resource management in DT-assisted vehicular networks.

    \item We propose a novel hybrid heuristic (HH) algorithm to efficiently address the vehicle assignment problem. Subsequently, an alternating optimization approach is employed to solve the non-convex joint optimization problem involving the semantic extraction ratio, beamforming matrices, and computational resource allocation. The optimal vehicle assignment enables efficient semantic communication and power-efficient semantic extraction, thereby reserving more power for signal transmission and DT modeling.
\end{enumerate}

The remainder of this paper is organized as follows: Section~II describes the system model of the NF-ISCSC system. Section~III formulates the performance indicators for the system. Section~IV describes the particle filter. Section~V focuses on the problem formulation and algorithm design. Simulation results are provided in Section~VI, and conclusions are drawn in Section~VII. A list of main symbols is summarized in Table.~\ref{tab:symbols}.
\begin{table}[htbp]
    \centering
    \caption{List of Main Symbols}
    \label{tab:symbols}
    \renewcommand{\arraystretch}{1.5}
    \begin{tabular}{|c|c|}
    \hline
    \textbf{Symbol} & \textbf{Description}  \\
    \hline\hline
    $\mathbf{\xi}_{m,k,i}$ & Binary indicator for RSU-vehicle communication service \\
    \hline
    $\mathbf{w}_{m,k,i}$ & Communication beamforming vector \\
    \hline
    $\mathbf{r}_{m,k,i}$ & Sensing beamforming vector \\
    \hline
    $\Bar{\mathbf{x}}_{m,i}$ & Communication signal with semantic information \\
    \hline
    $\mathbf{s}_{m,i}$ & Sensing signal \\
    \hline
    $\mathbf{x}_{m,i}$ & Joint signal \\
    \hline
    $ \mathbf{R}_{\mathbf{x}_{m,i}}$ & Covariance matrix of the joint signal \\
    \hline
    $\boldsymbol{\Gamma}_{m,k,i}$ & Path-loss matrix \\
    \hline 
    $\mathbf{A}\left( \theta, d\right)$ & Steering matrix with angle $\theta$ and distance $d$ \\
    \hline
    $\mathbf{z}_{m,k,i}$ & Echo signal before matched filtering\\
    \hline 
    $\mathbf{\hat{z}}_{m,k,i}$ & Echo signal after matching time-delay and Doppler shift\\
    \hline
    $\S_{m,k,i}$ & Semantic transmission rate\\
    \hline
    $\rho_{m,k,i}$ & Semantic extraction ratio\\
    \hline
    $\mathbf{J}_{m,k,i}$ & Fisher information matrix\\ 
    \hline
    $D_{m,k,i}$ & Data size\\
    \hline
    $C_{m,k,i}$ & Required computational resource for a given data size\\
    \hline 
    $f_{m,k,i}$ & CPU frequency allocated by the RSU\\
    \hline
    \end{tabular}
\end{table}

\subsection*{List of Notations:}
Matrices and vectors are denoted by boldface uppercase (e.g., $\mathbf{X}$) and lowercase letters (e.g., $\mathbf{x}$), respectively, while scalars appear in regular font. The sets $\mathbb{C}$, $\mathbb{C}^n$, and $\mathbb{C}^{m \times n}$ represent complex numbers, $n$-dimensional complex vectors, and $m \times n$ complex matrices, respectively. Key matrix operations include the Hermitian transpose $(\cdot)^H$, trace $\Tr(\cdot)$, and rank $\text{rank}(\cdot)$ operators. The identity and zero matrices are written as $\mathbf{I}$ and $\mathbf{0}$, respectively. The symbol $\succeq$ indicates positive semi-definiteness, and $\mathcal{CN}(0,\sigma^2)$ represents a complex Gaussian distribution with zero mean and variance $\sigma^2$.

\section{System Model}

We consider the design of an ISCSC system in a MU-MIMO configuration. We define the set of RSUs as $\mathcal{M} = \{1, 2, \ldots, M\}$, where $M$ is the total number of RSUs in the system, and each RSU $m \in \mathcal{M}$ is equipped with a uniform linear array (ULA) consisting of $N_t$ transmit antennas. The RSUs are strategically positioned along either side of the road, aligned parallel to the road surface. 
Let $\mathcal{K} = \{1, 2, \ldots, K\}$ denote the set of vehicles to be served, where each vehicle $k \in \mathcal{K}$ is equipped with $N_r$ receive antennas. All vehicles are assumed to travel within the NF region of the RSUs.
Each RSU $m$ aims to track all vehicles and build a DT model, and only communicate with a subset of vehicles in the system. Similar to the models presented in \cite{liu2020radar, dong2022sensing}, we assume that the vehicles travel at a constant velocity in the same direction, such as on a highway. However, in contrast to \cite{liu2020radar, dong2022sensing}, our scenario involves vehicles traveling along a double-lane straight road parallel to the RSUs, e.g., vehicles on a highway. Note that in some works, such as \cite{ding2024joint}, it is assumed that each RSU detects a subset of vehicles. In our case, each RSU maintains an offline DT model that locally represents its surrounding environment. Accordingly, each RSU detects and tracks all vehicles within its coverage area to ensure comprehensive awareness of road and traffic conditions for accurate DT construction and updates. Detecting only a subset of vehicles would correspond to an online DT scenario, where multiple RSUs cooperatively sense and fuse partial results via edge or cloud platforms. While an online DT framework inherently involves components such as distributed learning, cooperative sensing, and edge computing, these aspects lie beyond the scope of this work and are therefore left for future investigation.

\begin{figure*}[!t]
    \centering
    \includegraphics[width=\linewidth]{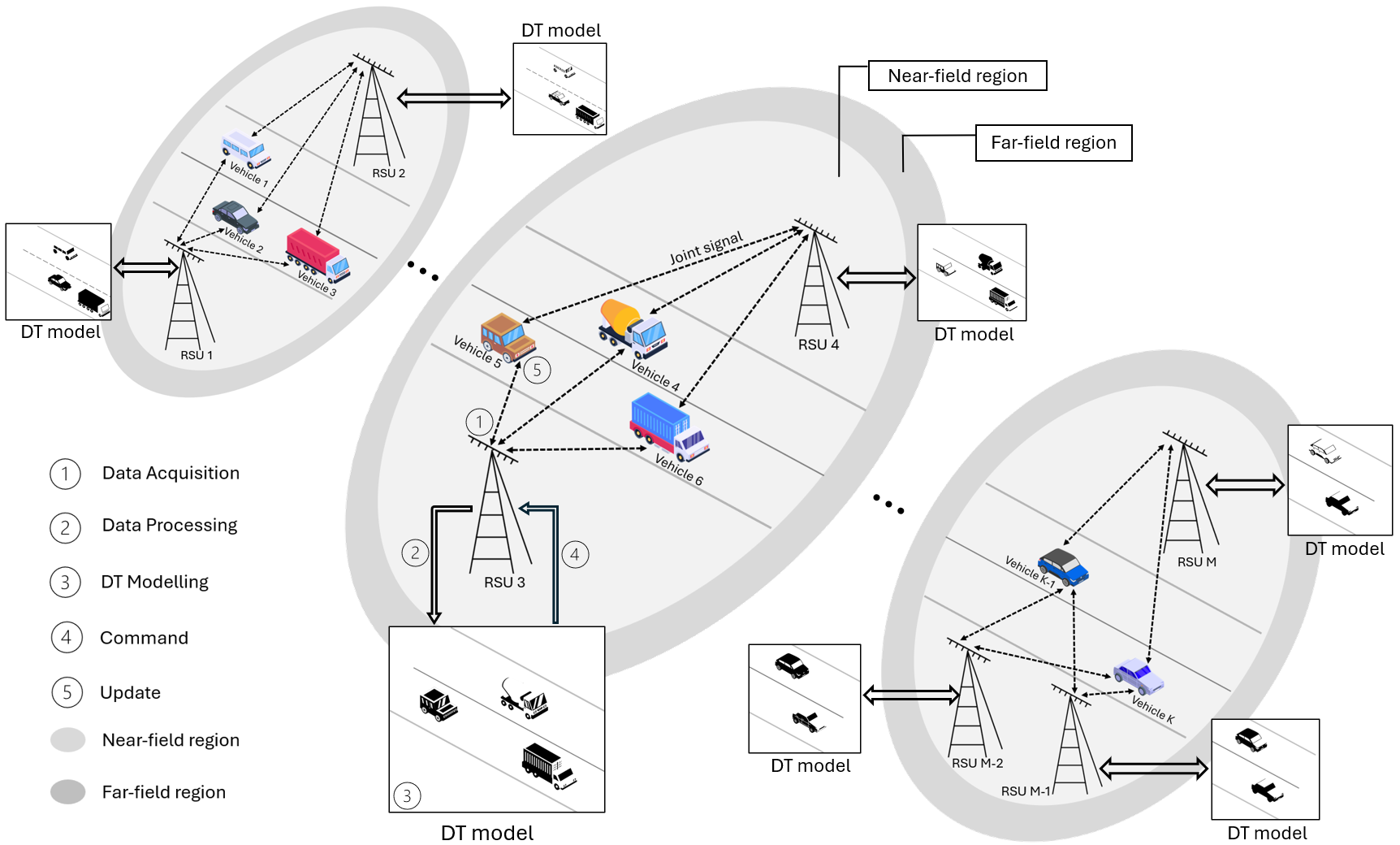}
    \caption{A system model of ISCSC for DT-assisted vehicular networks. Assuming an antenna aperture of 1 meter and RSUs operating at 50 GHz, the NF region extends to approximately 300 meters, covering the service range of a typical RSU. Each vehicle operates within the service coverage of the RSUs. For example, vehicles 1 to 3 move within the coverage area of RSU1 and RSU2. Each RSU aims to communicate with a subset of vehicles for road condition updates, while simultaneously detecting all vehicles on the road that are within its coverage to construct a DT model for real-time road condition prediction.} 
    \label{sys model}
\end{figure*}

The DT-enabled vehicular network, shown in Fig.~\ref{sys model}, operates as follows:
\begin{enumerate}
    \item {Data Acquisition:} Each RSU transmits joint signals to acquire information about surrounding vehicles and the environment, including location, velocity, and direction. 

    \item {Data Processing:} The acquired raw data is computed at the RSU using advanced signal processing algorithms, which may include filtering and object detection.

    \item {DT Modeling:} A DT model is constructed or updated based on the processed data. This virtual representation accurately reflects the physical environment and vehicular dynamics, enabling real-time simulation and behavior prediction.

    \item {Command:} Leveraging insights from the DT model, the RSU performs decision-making tasks such as route planning, collision avoidance, and speed regulation.

    \item {Update:} The RSU transmits control commands or warnings to vehicles, enabling adaptive behavior for enhanced safety and traffic efficiency. The control commands are jointly transmitted with the sensing signals.
    
\end{enumerate}

\subsection{Signal Model}
In the considered ISCSC system, communication and sensing signals are simultaneously transmitted from the RSUs to the vehicles through the use of beamformers. The communication beamforming vector for the $k$-th vehicle associated with the $m$-th RSU in time slot $i$ is defined as
\begin{equation}\label{eq5} 
\mathbf{\bar{w}}_{m,k,i} = \mathbf{\xi}_{m,k,i} \mathbf{w}_{m,k,i}, 
\end{equation}
where $\mathbf{w}_{m,k,i} \in \mathbb{C}^{N_t \times 1}$ represents the beamforming vector. The binary variable $\mathbf{\xi}_{m,k,i}$ indicates whether the $m$-th RSU serves the $k$-th vehicle in time slot $i$, with $\mathbf{\xi}_{m,k,i} = 1$ if the RSU is active and $\mathbf{\xi}_{m,k,i} = 0$ otherwise. Consequently, the transmitted signal from the $m$-th RSU is expressed as
\begin{equation} 
\mathbf{x}_{m,i} = \mathbf{\bar{W}}_{m,i} \Bar{\mathbf{x}}_{m,i} + \mathbf{R}_{m,i} \mathbf{s}_{m,i} , 
\end{equation}
where $\mathbf{\bar{W}}_{m,i} = [\mathbf{\bar{w}}_{m,1,i}, \mathbf{\bar{w}}_{m,2,i}, \ldots, \mathbf{\bar{w}}_{m,K,i}] \in \mathbb{C}^{N_t \times K}$ is the overall communication beamforming matrix, $\Bar{\mathbf{x}}_{m,i} = [\Bar{{x}}_{m,1,i}, \Bar{{x}}_{m,2,i}, \ldots, \Bar{{x}}_{m,K, i}] \in \mathbb{C}^{K \times 1}$ represents the semantic information. The semantic information $\Bar{x}_{m,k,i}$ can be obtained by applying a encoder function $f(\cdot)$, such as a Transformer, to the conventional message $c_{m,k,i}$, i.e., $\Bar{x}_{m,k,i} = f(c_{m,k,i})$.

Similarly, $\mathbf{R}_{m,i} = [\mathbf{r}_{m,1,i}, \mathbf{r}_{m,2,i}, \ldots, \mathbf{r}_{m,K,i}] \in \mathbb{C}^{N_t \times K}$ represents the sensing beamforming matrix and $\mathbf{s}_{m,i} = [{s}_{m,1,i}, {s}_{m,2,i}, \ldots, {s}_{m,K,i}] \in \mathbb{C}^{K \times 1}$ is the sensing signal. The design of the communication beamforming matrix $\mathbf{\bar{W}}_{m,i}$ and sensing beamforming matrix $\mathbf{R}_{m,i}$ are dynamically adjusted based on predictions of the vehicle's angle and distance from the RSU at the previous time slot $i-1$ (i.e., $\theta_{m,i|i-1}, d_{m,i|i-1}$). This predictive design helps ensure efficient and accurate transmission of both communication and sensing signals in the ISCSC system.

In addition, the covariance matrix for RSU $m$ of the transmitted signal is given by
\begin{equation} \label{covar}
    \mathbf{R}_{\mathbf{x}_{m,i}} = \mathbb{E} \left[\mathbf{x}_{m,i} \mathbf{x}_{m,i}^H \right] = \mathbf{\bar{W}}_{m,i} \mathbf{\bar{W}}_{m,i}^H + \mathbf{R}_{m,i}\mathbf{R}_{m,i}^H,
\end{equation}
with the assumption of $\mathbb{E} \left[\Bar{\mathbf{x}}_{m,i} \Bar{\mathbf{x}}_{m,i}^H \right] = \mathbf{I}$, $\mathbb{E} \left[{\mathbf{s}}_{m,i} {\mathbf{s}}_{m,i}^H \right] = \mathbf{I}$, and $\mathbb{E} \left[\Bar{\mathbf{x}}_{m,i} {\mathbf{s}}_{m,i}^H \right] = \mathbf{0}$.

\subsection{Communication Model}
In time slot $i$, the $k$-th vehicle receives signals from the $m$-th RSU. The corresponding formulation is characterized as follows:
\begin{equation}\label{eq1}
\begin{aligned}
    &\mathbf{y}_{m, k, i} = \underbrace{\boldsymbol{\Gamma}_{m,k,i} \mathbf{A}^H \left(\theta_{m,k,i}, d_{m,k,i}\right)  \mathbf{\bar{w}}_{m,k,i} \Bar{{x}}_{m,k,i}}_{\text{Desired signal}} +\mathbf{n}^c_{m,k,i}\\
    & \hspace{0.3cm} + \underbrace{\boldsymbol{\Gamma}_{m,k,i} \mathbf{A}^H \left(\theta_{m,k,i}, d_{m,k,i}\right)  \sum_{m=1}^M \sum_{k'=1, k' \neq k}^K \mathbf{\bar{w}}_{m,k',i} \Bar{{x}}_{m,k',i}}_{\text{Multi-RSU and multi-user communication interference}} \\
    & \hspace{0.3cm} + \underbrace{\boldsymbol{\Gamma}_{m,k,i} \mathbf{A}^H \left(\theta_{m,k,i}, d_{m,k,i}\right) \sum_{m=1}^M \sum_{k=1}^K \mathbf{r}_{m,k,i} {s}_{m,k,i} }_{\text{Multi-RSU and multi-user sensing interference}}, 
\end{aligned}
\end{equation}
where $\boldsymbol{\Gamma}_{m,k,i}$ represents the path-loss matrix, and $\mathbf{n}^c_{m,k,i} \sim \mathcal{CN}\left(0, \sigma^2_{c} \mathbf{I}\right)$ denotes the noise vector. The parameters $d_{m,k,i}$ and $\theta_{m,k,i}$ are the distance and angle of each vehicle measured with respect to the central reference points of the transmit and receive antenna arrays. respectively. The steering matrix is denoted by $\mathbf{A}\left(\theta_{m,k,i}, d_{m,k,i}\right) \in \mathbb{C}^{N_t \times N_r}$, and the formulations are given in \cite[Eq. (30)]{lin2024near}, which are shown below:
\scriptsize
\begin{equation}
\begin{aligned}
    &\mathbf{A}^H \left(\theta_{m,k,i}, d_{m,k,i} \right) =  \mathbf{a}^{R} \left( \theta_{m,k,i}, d_{m,k,i} \right) \\ 
    & \hspace{2cm} \cdot \left(\mathbf{a}^{T} \left(\theta_{m,k,i}, d_{m,k,i} \right) \right)^H  \odot \mathbf{H}\left(\theta_{m,k,i}, d_{m,k,i}\right),\\
    & \mathbf{a}^{T}\left(\theta_{m,k,i}, d_{m,k,i}\right) = e^{-j \frac{2 \pi}{\lambda} \left(-n_t d_t \cos \left(\theta_{m,k,i}\right) + \frac{n_t^2 d_t^2 \sin^2\left(\theta_{m,k,i}\right)}{2 d_{m,k,i}} \right)},\\
    & \mathbf{a}^{R}\left(\theta_{m,k,i}, d_{m,k,i}\right) = e^{-j \frac{2 \pi}{\lambda} \left(-n_r d_r \cos \left(\theta_{m,k,i}\right) + \frac{n_r^2 d_r^2 \sin^2 \left(\theta_{m,k,i}\right)}{2 d_{m,k,i}} \right)},\\
    & \mathbf{H}\left(\theta_{m,k,i}, d_{m,k,i}\right) = e^{-j \frac{2\pi}{\lambda d_{m,k,i}} \left(n_t d_t n_r d_r \sin^2 \left(\theta_{m,k,i}\right) \right)},
\end{aligned}
\end{equation}
\normalsize
where $n_t \in N_t$, $n_r \in N_r$, and $\lambda$ is the wavelength. The parameters $d_t$ and $d_r$ represent the spacing between adjacent antennas for the transmit and receive arrays, respectively. Each element of the path-loss matrix under a non-uniform spherical wave (NUSW) near-field channel model is given by \cite{liu2023near}
\begin{equation}
    \boldsymbol{\Gamma}_{m,k,i}[n_r, n_t]= \frac{1}{\sqrt{4 \pi \varphi_{m,k,i} }},
\end{equation}
with $\varphi_{m,k,i} = \biggl(d_{m,k,i}^2 + \left(n_t d_t + n_r d_r\right)^2 - 2 d_{m,k,i} \left(n_t d_t + n_r d_r\right) \cos \theta_{m,k,i} \biggr)$.

\subsection{Sensing Model}
The echo signal received by the $m$-th RSU encompasses information from all vehicles that it serves. Given that the RSU operates in a MIMO configuration, the channel correlation factor, which quantifies the similarity between signals received at different locations, asymptotically approaches zero, as demonstrated in \cite{liu2020radar, yuan2020bayesian, zhao2024modeling}. This implies that the reflected echoes from different vehicles exhibit negligible interference with one another. Consequently, the echo signal received by the $m$-th RSU corresponding to the $k$-th vehicle in time slot $i$ can be expressed as:
\begin{equation}\label{eq4}
\begin{aligned}
    \mathbf{z}_{m,k,i} &= \boldsymbol{\beta}_{m,k,i} e^{j 2\pi t \mu_{m,k,i}} \mathbf{A} \left(\theta_{m,k,i}, d_{m,k,i}\right)\\
    & \hspace{0.1cm} \cdot \mathbf{A}^H \left(\theta_{m,k,i}, d_{m,k,i}\right) \mathbf{x}_{m,k,i}(t - \tau_{m,k,i}) + \mathbf{n}^r_{m,k,i},  
 \end{aligned} 
\end{equation}
where $\boldsymbol{\beta}_{m,k,i}$ represents the round-trip path loss matrix, $\mathbf{x}_{m,k,i} \in \mathbb{C}^{N_t \times 1}$ is the transmitted signal for each vehicle, and $\mathbf{n}^r_{m,k,i} \sim \mathcal{CN}\left(0, \sigma^2_{r} \mathbf{I}\right)$ denotes the noise vector associated with the echo signal. Furthermore, $\mu_{m,k,i}$ is the Doppler frequency, which characterizes the frequency shift due to the relative motion between the vehicle and the RSU, while $\tau_{m,k,i}$ represents the time delay, capturing the propagation delay of the signal.

As stated in \cite{liu2020joint}, following the application of the matched filter, the time delay and Doppler frequency can be estimated via:
\footnotesize
\begin{equation}\label{matched filter}
    \{\hat{\tau}_{m,k,i}, \hat{\mu}_{m,k,i}\}= \arg \max_{\tau, \mu} \left|\int_0^{\Delta T} \mathbf{z}_{m,k,i} \mathbf{x}^*_{m,k,i}(t - \tau) e^{-j2 \pi\mu i} di \right|^2,
\end{equation}
\normalsize
where $\mathbf{x}^*_{m,k,i}\left(t-\tau\right)$ is the conjugate of the transmitted signal with a time shift $\tau$. The term $e^{-j2\pi \mu t}$ captures the Doppler shift in frequency, and the integration is performed over the observation interval $[0, \Delta T]$. The goal here is to maximize the correlation between the received signal and the delayed, frequency-shifted version of the transmitted signal, which allows for the accurate estimation of both the time delay $\tau_{m,k,i}$ and the Doppler frequency $\mu_{m,k,i}$.

Therefore, the distance $d_{m,k,i}$ and the velocity $v_{m,k,i}$ of the $k$-th vehicle in time slot $i$ can be estimated based on the time delay $\hat{\tau}_{m,k,i}$ and Doppler frequency $\hat{\mu}_{m,k,i}$ obtained from the matched filter. Specifically, the distance $d_{m,k,i}$ can be calculated as $\hat{d}_{m,k,i} = \frac{c \cdot \hat{\tau}_{m,k,i}}{2} + z_{\tau}$, where $c$ is the speed of light, and $z_{\tau}$ is the Gaussian noise with zero mean and variance of $\sigma^2_{2}$. Similarly, the velocity $v_{m,k,i}$ can be estimated from the Doppler frequency shift $\hat{\mu}_{m,k,i}$ using the expression
$\hat{v}_{m,k,i} = \frac{\lambda \hat{\mu}_{m,k,i} }{2} + z_{\mu}$, where $\lambda$ is the wavelength of the transmitted signal, and $z_{\mu}$ is the Gaussian noise with zero mean and variance of $\sigma^2_{3}$. 

\begin{remark}
    We assume that the matched filter outputs in \eqref{matched filter} are free from ambiguities, such as unresolved closely spaced angles, ranges, or Doppler shifts. In practice, the achievable angular, range, and velocity resolutions are determined by system parameters, such as the number of antennas (or aperture size), bandwidth, carrier frequency, waveforms, and the matched filter grid resolution. This assumption allows us to focus on the estimation level of the problem. 
\end{remark}

With the estimation of the time delay $\hat{\tau}_{m,k,i}$ and Doppler shift $\hat{\mu}_{m,k,i}$, the measurement of the angle $\theta_{m,k,i}$ and the round-trip path loss $\beta_{m,k,i}$ can be expressed as:
\footnotesize
\begin{equation}
    \hat{\mathbf{z}}_{m,k,i} = \boldsymbol{\beta}_{m,k,i} \mathbf{A}\left(\theta_{m,k,i}, d_{m,k,i} \right) \mathbf{A}^H \left(\theta_{m,k,i}, d_{m,k,i} \right) \mathbf{x}_{m,k,i}+ \mathbf{z}_{\theta},
\end{equation}
\normalsize
where $\mathbf{z}_\theta \sim \mathcal{CN}\left(0,\sigma^2_{1} \mathbf{I}\right)$ denotes the complex Gaussian measurement noise. Based on $ \hat{\mathbf{z}}_{m,k,i}$, the estimation of the angle $\theta_{m,k,i}$ can be achieved through high-resolution angle estimation techniques such as the MUSIC (MUltiple Signal Classification) algorithm, which is well-suited for separating multiple signal sources and accurately estimating the angle of arrival (AoA) by exploiting the eigenstructure of the covariance matrix of the received signals. Similarly, the round-trip path loss $\beta_{m,k,i}$ can be estimated using advanced algorithms like the Angle and Phase Estimation (APES) method \cite{liu2020joint}, which jointly estimates both the AoA and path loss by leveraging the phase and amplitude information of the received signal.

\subsection{Kinematic Model}
The kinematic model characterizes the temporal correlation between successive samples in the time domain, capturing the dynamic evolution of the vehicles. For vehicle $k$ served by RSU $m$ in time slot $i$, the state model is formulated as follows \cite{liu2020radar}:
\footnotesize
\begin{equation}
    \begin{aligned}
        & \theta_{m,k,i} = \theta_{m,k,i-1} + d^{-1}_{m,k,i-1} v_{m,k,i-1} \Delta T \sin\theta_{m,k,i-1} + u_\theta,\\
        & d_{m,k,i} = d_{m,k,i-1} - v_{m,k,i-1} \Delta T \cos\theta_{m,k,i-1} + u_d,\\
        & v_{m,k,i} = v_{m,k,i-1} + u_v,\\
        & \beta_{m,k,i} = \beta_{m,k,i-1} \left(1 + d^{-1}_{m,k,i-1} v_{m,k,i-1} \Delta T \cos\theta_{m,k,i-1} \right) + u_\beta, 
    \end{aligned}
\end{equation}
\normalsize
where $\mathbf{q}_{m,k,i} = [\theta_{m,k,i}, d_{m,k,i}, v_{m,k,i}, \beta_{m,k,i}]$ represents the state vector of vehicle $k$ served by RSU $m$ in time slot $i$, i.e., angle, distance, velocity, and path loss. $\Delta T$ is the length of a time slot. The term  $\mathbf{u}_i = [u_\theta, u_d, u_v, u_\beta]$ represents the state process noise.
By denoting the measured parameters as $\mathbf{r}_{m,k,i} = [\hat{\mathbf{z}}_{m,k,i},\hat{d}_{m,k,i}, \hat{v}_{m,k,i}]$ and the measurement noise as $\mathbf{z}_i = [\mathbf{z}_\theta, z_\tau, z_\mu]$, we can summarize the state model and the measurement model as follows:
\begin{equation}\label{eq55}
    \begin{cases}
      \text{State model:} & \mathbf{q}_{m,k,i} = \mathbf{g}_1\left(\mathbf{q}_{m,k,i-1}\right) + \mathbf{u}_i,\\
      \text{Measurement model:} & \mathbf{m}_{m,k,i} = \mathbf{g}_2\left(\mathbf{q}_{m,k,i}\right) + \mathbf{z}_i,
    \end{cases}
\end{equation}
where $\mathbf{g}_1\left(\cdot\right)$ is the state transition function and $\mathbf{g}_2\left(\cdot\right)$ is the measurement function. As $\mathbf{u}_i$ and $\mathbf{z}_i$ are noise vectors with zero-mean Gaussian distribution, their covariance matrices can be formulated by
\begin{equation}\label{eq56}
       \mathbf{Q}_1 = \diag \left(\sigma^2_\theta, \sigma^2_d, \sigma^2_v, \sigma_\beta^2 \right),\; \mathbf{Q}_2 = \diag \left(\sigma^2_1 \mathbf{I}, \sigma^2_{2}, \sigma^2_{3} \right),
\end{equation}
where the formulas for calculating $\sigma^2_{1}$, $\sigma^2_{2}$ and $\sigma^2_{3}$ are given in \cite[Eq. (24)]{liu2020radar}.

\section{Performance Measures}

In this section, we outline the performance measures used to evaluate the sensing, computing, and communication capabilities of the proposed ISCSC system.

\subsection{Semantic Communication}

The semantic transmission rate refers to the number of bits successfully received by the vehicle after extracting semantic information from the received signal. The mathematical formulation is shown in \eqref{eq12}, as outlined in \cite{yang2024secure, zhao2025compression, 11031224}.

\begin{table*}
\centering
\begin{minipage}{1\textwidth}
    \begin{align} \label{eq12}
    &S_{m,k,i} = \frac{\iota}{\rho_{m,k,i}} \log \left|\mathbf{I} + \frac{ \mathbf{H}_{m,k,i}^H \xi_{m,k,i}\mathbf{W}_{m,k,i} \mathbf{H}_{m,k,i}}{\mathbf{H}_{m,k,i}^H \left(\sum_{m=1}^M \sum^K_{k'=1, k' \neq k} \xi_{m,k',i} \mathbf{W}_{m,k',i} + \sum_{m=1}^M \sum_{k=1}^K \mathbf{R}_{m,k,i} \right) \mathbf{H}_{m,k,i}+ \sigma^2_{c} \mathbf{I}} \right|,
    \end{align}
\medskip
\hrule
\end{minipage}
\end{table*}

In \eqref{eq12}, the parameter $ 0\leq \rho_{m,k,i} \leq 1$ represents the semantic extraction ratio, and $\iota$ is a scalar value converting the word-to-bit ratio. Additionally, $\mathbf{H}_{m,k,i} = \boldsymbol{\Gamma}_{m,k,i} \mathbf{A} \left(\theta_{m,k,i}, d_{m,k,i} \right)$, $\mathbf{W}_{m,k,i} = \mathbf{w}_{m,k,i} \mathbf{w}_{m,k,i}^H$, $\mathbf{W}_{m,k',i} =  \mathbf{w}_{m,k',i} \mathbf{w}_{m,k',i}^H$, and $\mathbf{R}_{m,k,i} = \mathbf{r}_{m,k,i} \mathbf{r}_{m,k,i}^H$.

The lower bound of the semantic extraction ratio has been derived in \cite{yang2024secure}, which can be formulated as
\begin{equation}\label{eq14}
    \rho_{m,k,i} \geq \frac{1}{1 - \ln Q + \sum_{g=1}^G w_{g, m, k, i} \log p_{g, m, k, i} },
\end{equation}
where $Q$ represents the global lower bound of all the individual Bilingual Evaluation Understudy (BLEU) scores. The BLEU score evaluates how closely the reconstructed message, obtained from the received semantic information, matches the original message. Additionally, $w_{g,m,k,i}$ denotes the weight assigned to the $g$-grams, where $G$ is the total number of $g$-grams required to represent a sentence. The precision score $p_{g,m,k,i}$ is vehicle-specific and quantifies the accuracy of the message recovered by vehicle $k$ in time slot $i$.

A lower semantic extraction ratio results in a higher semantic transmission rate but also increases power consumption for semantic extraction. Under a limited power budget, this reduces the power available for signal transmission and DT modeling (to be discussed in subsequent sections), i.e., both $P^{\text{C\&S}}_{m,i}$ and $P_{m,k,i}^{\text{DT}}$ decrease, which adversely affects signal quality and the accuracy of the DT model.

\subsection{Sensing}

The mean square error (MSE) is frequently employed as a key metric for evaluating sensing performance, which compares the estimated parameter with its true value. However, deriving a closed-form expression for MSE can be very complex and computationally demanding, as highlighted in \cite{bekkerman2006target}. 

To address this challenge, we utilize the Cramér-Rao bound (CRB), which provides a theoretical lower bound on the variance of any unbiased estimator.
We define the parameters to be estimated as $\Psi_{m,k,i} = [d_{m,k,i}, \theta_{m,k,i}, \beta_{m,k,i}]$, representing the distance, angle, and round-trip path loss for vehicle $k$ served by RSU $m$ in time slot $i$, respectively. The Fisher information matrix (FIM), which quantifies the amount of information that the observed data carries about the parameter vector $\Psi_{m,k,i}$, is expressed as follows:
\begin{equation}\label{eq21}
\begin{aligned}
    \mathbf{J}_{m,k,i} &= 
    \begin{bmatrix}
        {J}_{d_{m,k,i} d_{m,k,i}} & {J}_{d_{m,k,i} \theta_{m,k,i}} &  \mathbf{J}_{d_{m,k,i} \beta_{m,k,i}}\\
        {J}_{\theta_{m,k,i} d_{m,k,i}} & {J}_{\theta_{m,k,i} \theta_{m,k,i}} &  \mathbf{J}_{\theta_{m,k,i} \beta_{m,k,i}}\\
        \mathbf{J}_{\beta_{m,k,i} d_{m,k,i}} & \mathbf{J}_{\beta_{m,k,i} \theta_{m,k,i} } &  \mathbf{J}_{\beta_{m,k,i} \beta_{m,k,i}}\\
    \end{bmatrix}\\
    & = \begin{bmatrix}
        \mathbf{J}_{m,k,i,11} & \mathbf{J}_{m,k,i,12} \\
        \mathbf{J}_{m,k,i,12}^T & \mathbf{J}_{m,k,i,22}\\
        \end{bmatrix}.
\end{aligned}
\end{equation}

For any $a,b \in \{\theta_{m,k,i}, d_{m,k,i}\}$, the corresponding FIM elements can be computed as follows:
\begin{equation}\label{eq22}
    {J}_{a, b} = \frac{2T|\beta_{m,k,i}|^2}{\sigma^2_{z}} \Tr \left(\dot{\mathbf{B}}_{b} \mathbf{R}_{\mathbf{x}_{m,i}} \dot{\mathbf{B}}_{a}^H \right),
\end{equation}
\begin{equation}\label{eq23}
    \mathbf{J}_{a \beta_{m,k,i}} = \frac{2T \beta_{m,k,i}^*}{\sigma^2_{z}} \Re \left\{ \Tr \left(\mathbf{B}_{m,k,i} \mathbf{R}_{\mathbf{x}_{m,i}} \dot{\mathbf{B}}^H_{a} \right)\right\}[1 \; j],
\end{equation}
\begin{equation}\label{eq24}
    \mathbf{J}_{\beta_{m,k,i} \beta_{m,k,i}} = \frac{2T}{\sigma^2_{z}} \Tr \left(\mathbf{B}_{m,k,i} \mathbf{R}_{\mathbf{x}_{m,i}} \mathbf{B}_{m,k,i}^H \right) \mathbf{I},
\end{equation}
where $\mathbf{B}_{m,k,i} = \mathbf{A} \left(\theta_{m,k,i}, d_{m,k,i} \right) \mathbf{A}^H \left(\theta_{m,k,i}, s_{m,k,i}\right)$. We further denote $\Dot{\mathbf{B}}_{\theta_{m,k,i}} = \frac{\partial \mathbf{B}_{m,k,i}}{\partial \theta_{m,k,i}}$ and $\Dot{\mathbf{B}}_{r_{m,k,i}} = \frac{\partial \mathbf{B}_{m,k,i}}{\partial d_{m,k,i}}$ representing the partial derivatives of $\mathbf{B}_{m,k,i}$ with respect to the angle $\theta_{m,k,i}$ and distance $d_{m,k,i}$, respectively.

Using the FIM elements derived, the CRB for $\theta_{m,k,i}$ and $d_{m,k,i}$, which are the parameters of primary interest, can be expressed as follows:
\begin{equation}\label{eq25}
     \text{CRB}\left(d_{m,k,i}\right) = {\mathbf{J}_{m,k,i}^{-1}}_{[1,1]}, \; \text{CRB}\left(\theta_{m,k,i}\right) = {\mathbf{J}_{m,k,i}^{-1}}_{[2,2]},
\end{equation}
where $\mathbf{A}_{[i,j]}$ means the element of matrix $\mathbf{A}$ in row $i$ and column $j$, and $\mathbf{J}_{m,k,i}^{-1} =\left(\mathbf{J}_{m,k,i,11} - \mathbf{J}_{m,k,i,12} \mathbf{J}^{-1}_{m,k,i,22} \mathbf{J}_{m,k,i, 12}^T \right)^{-1}$.

\subsection{Computing for Semantic Communication and Sensing}

Extracting semantic information from traditional messages predominantly depends on advanced machine learning techniques, which introduce significant computational overhead. Consequently, it is essential to account for computational power as an integral part of the overall transmission power budget. As detailed in \cite{yang2024secure,zhao2025compression}, the computational power consumption is modeled using a natural logarithmic function to capture the relationship between the computational complexity and power requirements. The formulation is given by:
\begin{equation}\label{eq18}
    P^{\text{Comp}}_{m,i} =  -F\sum_{k=1}^K \mathbf{\xi}_{m,k,i} \ln\left(\rho_{m,k,i}\right),
\end{equation}
where $F$ is a coefficient that converts a magnitude to its power.

On the other hand, the communication and sensing power consumption at the $m$-th RSU is given by
\begin{equation}\label{eq19}
    P^{\text{C\&S}}_{m,i} = \Tr\left( \mathbf{\bar{W}}_{m,i} \mathbf{\bar{W}}_{m,i}^H + \mathbf{R}_{m,i}\mathbf{R}_{m,i}^H \right).
\end{equation}

\subsection{Computing for Digital Twin}

In DT-assisted vehicular networks, real-time updates are critical for maintaining accurate and timely digital representations of vehicles and their surrounding environment. To achieve this, RSUs must efficiently process large volumes of sensed data. Given the limited computational resources available at the RSUs, it becomes essential to estimate and manage the computational workload required for updating DT models. This ensures an optimal balance between processing latency and energy consumption, thereby supporting reliable and scalable DT operations within stringent real-time constraints. 
The computation task assigned to the $m$-th RSU for vehicle $k$ is denoted by the tuple $\left(D_{m,k,i}, C_{m,k,i}\right)$, where $D_{m,k,i}$ represents the data size in bits, and $C_{m,k,i}$ is the required computational resource in terms of CPU cycles per bit. As shown in various studies, including \cite{liu2021digital, huang2023mobile, van2022edge}, the processing latency for such a task can be expressed as:
\begin{equation}
    T_{m,k,i} =  \frac{C_{m,k,i} D_{m,k,i}}{f_{m,k,i}} + \Delta T_{m, k, i},
\end{equation}
where $f_{m,k,i}$ represents the CPU frequency allocated by the RSU for constructing the DT model, while $\Delta T_{m, k, i}$ represents the additional processing latency incurred due to errors in the collected data. 
Despite its significance, existing research has not proposed a method for estimating $\Delta T_{m, k, i}$. This limitation arises primarily because most prior works focus on a single time slot or assume a static scenario, under which it is reasonable to treat $\Delta T_{m, k, i}$ as a constant. However, in dynamic environments, this assumption no longer holds, and $\Delta T_{m, k, i}$ must be explicitly considered. To bridge this gap, we introduce a novel approach for its estimation.

The additional processing latency incurred due to errors in the data to build a DT model can be estimated as:
\begin{equation}\label{dt_RCRB}
    \Delta T_{m, k, i} = \frac{\mathcal{L}}{f_{m,k,i}},
\end{equation}
where the workload $\mathcal{L}$ is given by:
\begin{equation}\label{dt_RCRB2}
    \mathcal{L} =\nu_1 f_1 \left( \text{RCRB} \left(d_{m,k,i-1} \right) \right) + \nu_2 f_2 \left( \text{RCRB} \left(\theta_{m,k,i-1} \right) \right) + \nu_3.
\end{equation}

In \eqref{dt_RCRB2}, $\mathcal{L}$ represents the extra computational workload required for DT modeling, $\nu_1$ and $\nu_2$ are scaling coefficients that translate sensing errors (in distance and angle) into computational requirements, and $\nu_3$ accounts for additional processing overhead that are not explicitly modeled in this paper (e.g., temperature variations). The terms $f_1\left(\cdot\right)$ and $f_2\left(\cdot\right)$ are functions of the RCRB values for distance $\left(d_{m,k,i-1}\right)$ and angle $\left(\theta_{m,k,i-1}\right)$, respectively, from the previous time slot.

\begin{remark}
    Equation \eqref{dt_RCRB} establishes a direct relationship between sensing accuracy and computational cost.
    Specifically, as the sensing performance for angle or distance degrades, the corresponding RCRB values increase. This increase indicates a larger discrepancy between the true and detected values, resulting in a higher computational load $\mathcal{L}$, which in turn leads to greater processing time. Given that the total latency $T_{m,k,i}$ is usually constrained by the maximum allowable latency $T^{\text{max}}$, the CPU frequency $f_{m,k,i}$ must be increased to meet this latency requirement. However, this adjustment leads to increased power consumption. Alternatively, improving the sensing performance can help minimize the computational workload $\mathcal{L}$, thereby reducing the need for higher CPU frequencies and decreasing power consumption.
\end{remark}

The power consumption for executing such a task can be formulated as \cite{lu2020communication, do2022digital, dong2019deep}:
\begin{equation}
    P_{m,k,i}^{\text{DT}} = \kappa \left(f_{m,k,i}\right)^3 C_{m,k,i},
\end{equation}
where $\kappa$ denotes the energy-efficiency coefficient, which is dependent on the CPU design.

\section{Particle Filter}

This paper adopts the particle filter (PF) due to its robust performance in noisy environments compared to the extended Kalman filter (EKF) \cite{djuric2003particle}. PF is a sequential Monte Carlo method used to estimate the \textit{a posterior} distribution of a system's state, offering greater flexibility in handling non-linear and non-Gaussian models. 

\subsection{The Bayesian Filters}

The core objective in any Bayesian filtering method, including PF and EKF, is to estimate the \textit{a posterior} distribution of the vehicle state given all observations up to the current time, i.e., $p\left(\mathbf{q}_{m,k,i} | \mathbf{m}_{m,k,1:i} \right)$.

Given the \textit{a prior} distribution of the initial state $p\left(\mathbf{q}_{m,k,0}\right) $, the state transition distribution $p\left(\mathbf{q}_{m,k,i} | \mathbf{q}_{m,k,i-1}\right) $, and the likelihood function $ p\left(\mathbf{m}_{m,k,i} | \mathbf{q}_{m,k,i}\right) $, we can update the \textit{a prior} distribution of the state at each time slot $i$ by
\begin{equation}\label{prior}
\begin{aligned}
    &p\left( \mathbf{q}_{m,k,i} | \mathbf{m}_{m,k,1:i-1} \right) \\
    &= \int p\left( \mathbf{q}_{m,k,i} | \mathbf{q}_{m,k,i-1} \right) p\left( \mathbf{q}_{m,k,i-1} | \mathbf{m}_{m,k,1:i-1} \right) d \mathbf{q}_{m,k,i-1},
\end{aligned}
\end{equation}
where the initial or the previous \textit{a posterior} distribution $p\left(\mathbf{q}_{m,k,i-1} | \mathbf{m}_{m,k,1:i-1}\right) $ is assumed to be known or already known.

Once the new observation $\mathbf{m}_{m,k,i}$ is received, the \textit{a posterior} distribution is updated using the Bayes' rule:
\begin{equation}\label{eq27?}
\begin{aligned}
        & p\left( \mathbf{q}_{m,k,i} | \mathbf{m}_{m,k,1:i} \right) \\
        &= \frac{p\left( \mathbf{m}_{m,k,i} | \mathbf{q}_{m,k,i} \right) p\left( \mathbf{q}_{m,k,i} | \mathbf{m}_{m,k,1:i-1} \right)}{\int p\left( \mathbf{m}_{m,k,i} | \mathbf{q}_{m,k,i} \right) p\left( \mathbf{q}_{m,k,i} | \mathbf{m}_{m,k,1:i-1} \right) d \mathbf{q}_{m,k,i}}.
\end{aligned}
\end{equation}

The integrals in \eqref{prior} and \eqref{eq27?} can be computationally intractable for nonlinear or high-dimensional systems, leading to difficulties in estimating $ p\left( \mathbf{q}_{m,k,i} | \mathbf{m}_{m,k,1:i} \right)$. Traditional methods, such as the EKF, approximate these integrals by linearizing the models and assuming Gaussian distributions, which may be inaccurate in many practical scenarios \cite{ristic2003beyond}.

\subsection{Particle Filter Implementation}

The state parameter \(\mathbf{q}_{m,k,i}\) can be estimated via
\begin{equation}\label{PT estimation}
    \hat{\mathbf{q}}_{m,k,i} = \int \mathbf{q}_{m,k,i} \, p(\mathbf{q}_{m,k,i} \mid \mathbf{m}_{m,k,1:i}) \, d \mathbf{q}_{m,k,i}.
\end{equation}

However, the \textit{a posterior} distribution $p\left(\mathbf{q}_{m,k,i} | \mathbf{m}_{m,k, 1:i}\right)$ is unknown or hard to compute. The PF approximates this distribution by a weighted set of particles \cite{djuric2003particle, chen2003bayesian, arulampalam2002tutorial}:
\begin{equation}
    p(\mathbf{q}_{m,k,i} \mid \mathbf{m}_{m,k,1:i}) \approx \sum_{n=1}^{N_s} \tilde{w}_{m,k,i}^n \, \delta(\mathbf{q}_{m,k,i} - \mathbf{q}_{m,k,i}^n),
\end{equation}
where $\delta\left(\cdot\right) $ denotes the Dirac delta function, $N_s$ is the number of particles, and $w_{m,k,i}^n$ is the weight associated with the $n$-th particle. Hence, \eqref{PT estimation} becomes
\begin{equation}\label{state approxi}
    \hat{\mathbf{q}}_{m,k,i} \approx \sum_{n=1}^{N_s} \tilde{w}_{m,k,i}^n \mathbf{q}_{m,k,i}^n.
\end{equation}

Each particle is propagated forward using a probabilistic model, often based on the state transition prior:
\begin{equation}
    \mathbf{q}_{m,k,i}^n \sim p \left(\mathbf{q}_{m,k,i} | \mathbf{q}_{m,k,i-1}^n \right).
\end{equation}

Then, the particle weights are updated using the likelihood of the current observation conditioned on the particle’s state:
\begin{equation} \label{weight_update}
    w_{m,k,i}^n = w_{m,k,i-1}^n \, p \left(\mathbf{m}_{m,k,i} | \mathbf{q}^n_{m,k,i} \right),
\end{equation}
and the unnormalized weights are subsequently normalized as:
\begin{equation} \label{norm weight}
    \tilde{w}_{m,k,i}^n = \frac{w_{m,k,i}^n}{\sum_{n=1}^{N_s} w_{m,k,i}^n}.
\end{equation}

The procedure for implementing the PF is summarized in \textbf{Algorithm \ref{alg:pf}}.

\begin{algorithm}[t]
    \caption{Particle Filter}
    \label{alg:pf}
    \begin{algorithmic}[1]
        \STATE Initialize particles $\mathbf{q}_{m,k,0}^n$ by drawing from the \textit{a prior} distribution $p\left(\mathbf{q}_{m,k,0}\right)$.
        \STATE Set initial weights $w_{m,k,0}^n = \frac{1}{N_s}$.
        \FOR{each time slot $i$}
            \FOR{each particle $n = 1, \dots, N_s$}
                \STATE Sample $\mathbf{q}_{m,k,i}^n$ from $p\left(\mathbf{q}_{m,k,i} | \mathbf{q}_{m,k,i-1}^n\right)$.
                \STATE Compute weight $w_{m,k,i}^n$ using \eqref{weight_update}.
            \ENDFOR
            \STATE Normalize the weights by applying \eqref{norm weight}.
            \STATE Resample the particles via the multinomial method based on $\tilde{w}_{m,k,i}^n$ to prevent particle depletion.
            \vspace{0.1cm}
            \STATE Update the prediction of $\mathbf{q}_{m,k,i}$ by \eqref{state approxi}.
        \ENDFOR
    \end{algorithmic}
\end{algorithm}

\section{ISCSC Design for DT-enabled Vehicular Networks}

\subsection{Problem Formulation}

The design objective is to maximize the overall system semantic transmission rate while minimizing the overall CRB, ensuring both high communication efficiency and precise sensing performance. These dual objectives can be formulated into the following optimization problem:
\begin{subequations} \label{opt1}
\begin{align}
    \min_{\chi} \quad &   \sum_{m=1}^M \sum_{k=1}^{K} \varepsilon  \left(- \eta_{S_{m,k,i}} \right) + \left(1-\varepsilon\right) \left(\eta_{\theta_{m,k,i}} + \eta_{d_{m,k,i}}\right) \label{opt1a}\\
    \text{s.t.} \quad & -S_{m,k,i} \leq -\eta_{S_{m,k,i}}, \ \forall k, \forall m, \label{opt1b}\\
    & \text{CRB}\left(\theta_{m,k,i}\right) \leq \eta_{\theta_{m,k,i}}, \ \forall k, \forall m, \label{opt1h}\\
    & \text{CRB}\left(d_{m,k,i}\right) \leq \eta_{d_{m,k,i}}, \ \forall k, \forall m, \label{opt1i}\\
    & P^{\text{C\&S}}_{m,i} + P^{\text{Comp}}_{m,i}  + \sum_{k=1}^K P_{m,k, i}^{\text{DT}} \leq P_t, \ \forall m, \label{opt1c}\\
    & \sum_{m=1}^M \mathbf{\xi}_{m,k,i} = 1, \ \forall k, \label{opt1d}\\
    & \max_{k \in \mathcal{K}}  T_{m,k,i} \leq T^{\text{max}}, \ \forall m, \label{opt1e}\\
    & \sum_{k=1}^K f_{m,k,i} \leq F^{\text{max}}, \ \forall m, \label{opt1f}\\
    & p_{LB} \leq \rho_{m,k,i} \leq 1, \ \forall k, \forall m, \label{opt1g}
\end{align}
\end{subequations}
where $\varepsilon$ is the weight, and the optimization variable $\chi$ consists of the set $\{\mathbf{W}_{m,k,i}~\succeq 0, \mathbf{R}_{m,k,i}~\succeq 0,\mathbf{\xi}_{m,k,i}, \eta_{S_{m,k,i}},\\ \eta_{\theta_{m,k,i}}, \eta_{d_{m,k,i}}, \rho_{m,k,i}, f_{m,k,i} \}$.

The constraint \eqref{opt1c} enforces that the total power consumed for semantic extraction, semantic communication, sensing, and DT model construction must remain within the transmission power budget $P_t$. Moreover, \eqref{opt1d} ensures that each vehicle $k$ is exclusively served by one RSU in any time slot $i$. The constraint \eqref{opt1e} guarantees the worst-case latency for processing tasks at RSU $m$ does not exceed the maximum allowable delay $T^{\text{max}}$, while constraint \eqref{opt1f} restricts the total CPU frequency allocation for each RSU to stay below the maximum available frequency $F^{\text{max}}$. Finally, \eqref{opt1g} ensures that the semantic extraction ratio $\rho_{m,k,i}$ remains within specified bounds, and $\rho_{LB}$ is given in \eqref{eq14}.

It is important to note that the rank-one constraints for the beamforming matrices have been eliminated in this formulation. A rank-one solution can be reconstructed after solving the optimization problem by Gaussian randomization.

\subsection{Problem Transformation}

The optimization problem \eqref{opt1} is non-convex, primarily due to the non-convex nature of the constraints \eqref{opt1b}, \eqref{opt1h}, and \eqref{opt1i}. Therefore, we present techniques for reformulating these elements into equivalent convex or concave representations, enabling more efficient and tractable optimization.

First, we address the non-convex constraint in \eqref{opt1b}. The transmission rate $S_{m,k,i}$ in \eqref{eq12} can be reformulated as shown in \eqref{eq31?}, which can be found at the top of the next page. 
\begin{table*}
\footnotesize
\centering
\begin{minipage}{1\textwidth}
    \begin{align} \label{eq31?}
        -S_{m,k,i} &= \frac{\iota}{\rho_{m,k,i}} \left( \underbrace{\log \left|\mathbf{I} + \frac{\mathbf{H}_{m,k,i}^H \left( \sum_{m=1}^M \sum_{k'=1, k' \neq k}^K \xi_{m,k',i}\mathbf{W}_{m,k',i} + \sum_{m=1}^M \sum_{k=1}^K \mathbf{R}_{m,k,i}  \right) \mathbf{H}_{m,k,i}}{\sigma_{c}^2}\right|}_{\text{The first term}} \nonumber \right. \\
        &\left. \hspace{2cm} \underbrace{- \log \left|\mathbf{I} + \frac{\mathbf{H}_{m,k,i}^H \left( \xi_{m,k,i}\mathbf{W}_{m,k,i} + \sum_{m=1}^M\sum^K_{k'=1, k' \neq k} \xi_{m,k',i}\mathbf{W}_{m,k',i} + \sum_{m=1}^M\sum_{k=1}^K \mathbf{R}_{m,k,i} \right) \mathbf{H}_{m,k,i}}{\sigma_{c}^2}\right|}_{\text{The second term}} \right).
    \end{align}
\medskip
\hrule
\end{minipage}
\end{table*}

\begin{lemma}[\hspace{-0.02mm}\cite{christensen2008weighted}]
If $\mathbf{E} \in \mathbb{C}^{N \times N}$ is a Hermitian positive definite matrix, the following equation holds by introducing a supplementary variable $\mathbf{S}$:
\begin{equation}
    \ln \left|\mathbf{E}^{-1}\right|=\min_{\mathbf{A} \succeq 0, \mathbf{A} \in \mathbb{C}^{N \times N}} \Tr\left(\mathbf{A} \mathbf{E}\right) - \ln|\mathbf{A}| - N.
\end{equation}
\end{lemma}

By applying \textbf{Lemma 1}, the second term in \eqref{eq31?} can be reformulated into a convex form, as shown in \eqref{second term} at the top of the next page. In \eqref{second term}, $\mathbf{A}_{m,k,i} \in \mathbb{C}^{N_r \times N_r}$ and $\mathbf{A}_{m,k,i} \succeq 0$. However, the first term in \eqref{eq31?} is still non-convex.

\begin{table*}
\footnotesize
\centering
\begin{minipage}{1\textwidth}
    \begin{align}\label{second term}
         \Tr \left(\mathbf{A}_{m,k,i} \left(\mathbf{I} + \frac{\mathbf{H}_{m,k,i}^H \left( \xi_{m,k,i}\mathbf{W}_{m,k,i} + \sum_{m=1}^M\sum^K_{k'=1, k' \neq k} \xi_{m,k',i}\mathbf{W}_{m,k',i} + \sum_{m=1}^M\sum_{k=1}^K \mathbf{R}_{m,k,i} \right) \mathbf{H}_{m,k,i}}{\sigma_{c}^2}\right)\right) - \ln \left|\mathbf{A}_{m,k,i}\right| - N_r .
    \end{align}
\medskip
\hrule
\end{minipage}
\end{table*}   

\begin{lemma}[Low SNR approximation \cite{verdu2002spectral}]
\begin{equation}
    \log \left|\mathbf{I} + \frac{\mathbf{H}^H \mathbf{V} \mathbf{H}}{\sigma^2}\right| \approx \frac{1}{\sigma^2} \Tr\left(\mathbf{H}^H \mathbf{V} \mathbf{H}\right). 
\end{equation}
\end{lemma}

Since the first term in \eqref{eq31?} appears in the denominator of $S_{m,k,i}$, maximizing $S_{m,k,i}$ implicitly requires minimizing this term. Therefore, it is reasonable to assume that the first term in \eqref{eq31?} remains small. As such, we can use \textbf{Lemma 2} to approximate the first term in \eqref{eq31?}, which leads to the following result:
\begin{equation}\label{first term}
\begin{aligned}
    &\frac{1}{\sigma_{c}^2}\Tr \left(\mathbf{H}_{m,k,i}^H \left( \sum_{m=1}^M \sum_{k'=1, k' \neq k}^K \xi_{m,k',i}\mathbf{W}_{m,k',i} \right. \right. \nonumber \\
    & \left. \left. \hspace{0.5cm} + \sum_{m=1}^M \sum_{k=1}^K \mathbf{R}_{m,k,i}  \right) \mathbf{H}_{m,k,i} \right ).
\end{aligned}
\end{equation}

Combining \eqref{second term} and \eqref{first term}, we transform \eqref{eq31?} into the convex form as shown in \eqref{eq31transf} at the top of the next page.

\begin{remark}
    Lemma~1 introduces an auxiliary variable to turn $\ln|\mathbf{E}^{-1}|$ into a convex form, while Lemma~2 offers a low-SNR approximation of the logarithmic determinant when the signal power is small relative to noise. To ensure consistency between the two lemmas, the rate expression is reformulated as $\min -S_{m,k,i} = \log|\mathbf{A}^{-1}| + \log|\mathbf{B}|$, instead of directly maximizing $S_{m,k,i} = \log|\mathbf{A}| - \log|\mathbf{B}|$. This inversion allows Lemma~1 and Lemma~2 to be correctly applied to the high SNR term $\log|\mathbf{A}^{-1}|$ and the low SNR term $\log|\mathbf{B}|$, respectively.
\end{remark}

\begin{table*}
\centering
\begin{minipage}{1\textwidth}
    \begin{align} \label{eq31transf}
      &\frac{\iota}{\rho_{m,k,i}} \left( \frac{1}{\sigma_{c}^2}\Tr \left( \mathbf{H}_{m,k,i}^H \left( \sum_{m=1}^M \sum_{k'=1, k' \neq k}^K \xi_{m,k',i}\mathbf{W}_{m,k',i} + \sum_{m=1}^M \sum_{k=1}^K \mathbf{R}_{m,k,i}  \right) \mathbf{H}_{m,k,i} \right)  - \ln \left|\mathbf{A}_{m,k,i}\right| - N_r \nonumber \right.\\
      & \left.  \Tr \left( \mathbf{A}_{m,k,i} \left( \mathbf{I} + \frac{\mathbf{H}_{m,k,i}^H \left( \xi_{m,k,i}\mathbf{W}_{m,k,i} + \sum_{m=1}^M\sum^K_{k'=1, k' \neq k} \xi_{m,k',i}\mathbf{W}_{m,k',i} + \sum_{m=1}^M\sum_{k=1}^K \mathbf{R}_{m,k,i} \right) \mathbf{H}_{m,k,i}}{\sigma_{c}^2} \right) \right) \right) \leq -\eta_{S_{m,k,i}}, \forall k, \forall m.
    \end{align}
\medskip
\hrule
\end{minipage}
\end{table*}

Next, we address the non-convex constraints \eqref{opt1h} and \eqref{opt1i}. These two constraints can be combined as $\Tr \left( \text{CRB}\left(\mathbf{\Xi}_{m,k,i}\right)  \right) \leq \eta_{m,k,i}$, where $\mathbf{\Xi}_{m,k,i} = \begin{bmatrix} \theta_{m,k,i} & d_{m,k,i} \end{bmatrix}$. According to \cite{lyu2024crb}, minimizing $\Tr\left( \text{CRB} \left(\mathbf{\Xi}\right) \right) \leq \eta$ is equivalent to maximizing its upper bound $\Tr \left(\mathbf{\Omega}^{-1}\right)$, where $\mathbf{J}_{11} - \mathbf{J}_{12} \mathbf{J}^{-1}_{22} \mathbf{J}_{12}^T \succeq \mathbf{\Omega}$. Thus, by applying the Schur complement, we can replace the non-convex CRB constraints with the following convex constraints:
\begin{equation}\label{eq45}
    \begin{aligned}
        &\begin{bmatrix}
            \mathbf{J}_{m,k,i,11} - \mathbf{\Omega}_{m,k,i} & \mathbf{J}_{m,k,i,12} \\
            \mathbf{J}_{m,k,i,12}^T & \mathbf{J}_{m,k,i,22}
        \end{bmatrix} \succeq 0, \\
        &\Tr \left(\mathbf{\Omega}_{m,k,i} ^{-1}\right) \leq \eta_{m,k,i}, \mathbf{\Omega}_{m,k,i} \succeq 0, \; \forall m, \forall k.  
    \end{aligned}
\end{equation}

Finally, with these transformations, the original non-convex optimization problem \eqref{opt1} is reformulated as a convex one:
\begin{subequations} \label{opt2}
\begin{align}
    \min_{\mathbf{\chi}} \quad & \sum_{m=1}^M \sum_{k=1}^K  \varepsilon  \left(-\eta_{S_{m,k,i}}\right) + \left(1 - \varepsilon\right) \eta_{m,k,i} \label{opt2a}\\
    \text{s.t.} \quad & \eqref{opt1c}, \eqref{opt1d}, \eqref{opt1e}, \eqref{opt1f}, \eqref{opt1g}, \eqref{eq31transf} , \eqref{eq45},
\end{align}
\end{subequations}
where $\mathbf{\chi} = \{ \mathbf{W}_{m,k,i}, \mathbf{R}_{m,k,i}, \xi_{m,k,i}, \rho_{m,k,i}, f_{m,k,i}, \\ 
\mathbf{A}_{m,k,i}, \eta_{S_{m,k,i}}, \eta_{m,k,i}, \mathbf{\Omega}_{m,k,i}\}$.

\subsection{Algorithm Design}
To solve the optimization problem \eqref{opt2}, we decompose it into two sub-problems: the outer optimization problem focuses on vehicle assignment by optimizing $\xi_{m,k,i}$, while the inner optimization problem addresses the optimization of the remaining variables in $\chi$.

We begin by discussing the outer optimization problem. In time slot $i$, the vehicle assignment problem is represented by the binary variable $\xi_{m,k,i}$, which selects the optimal RSU to serve each vehicle. For example, $\xi_{2,1,10}$ means that in time slot 10, vehicle 1 is served by RSU 2. This assignment problem is inherently a binary optimization problem. To efficiently solve it, we employ an HH approach combining greedy and simulated annealing (SA) algorithms. The greedy algorithm initially assigns each vehicle to the nearest RSU based on distance, aiming to minimize the impact of path loss on communication performance and improve the precision of sensing results. This approach is computationally efficient, making it well-suited for real-time vehicular applications. However, the greedy algorithm is prone to sub-optimal solutions, particularly in scenarios with varying vehicle densities. To overcome this limitation, we enhance the solution by applying the SA algorithm, which helps escape local optima and explore a broader solution space. SA improves upon the initial greedy solution by introducing randomness into the search process and gradually refining the assignment over iterations, ultimately leading to a potentially better vehicle assignment plan. The pseudocode for the HH algorithm is outlined in \textbf{Algorithm \ref{alg HH}}, and the optimal vehicle assignment solution is denoted as $\xi_{\text{best}}$. This hybrid approach balances the simplicity and speed of the greedy algorithm with the robustness of SA, ensuring both real-time applicability and improved solution quality for vehicle assignment.

\begin{algorithm}
\caption{Hybrid Heuristic Algorithm for Vehicle Assignment}\label{alg HH}
\begin{algorithmic}[1]
\STATE Create a tabu list $\psi$.
\REPEAT
    \STATE Calculate the distance between each RSU $m$ and vehicle $k$, denoted by $d_{m,k,i}$.
    \STATE Assign each vehicle $k$ to the RSU with the minimum distance: $\xi_{m,k,i} = 1$ where $m = \arg \min_{m \in M} d_{m,k,i}$.
\UNTIL{all vehicles are assigned to an RSU, i.e., constraint \eqref{opt1d} is satisfied.}
\IF{$\xi_{m,i} \notin \psi$} 
    \STATE The initial feasible solution, $\xi_{m,i}^{\text{init}}$, is obtained.
\ELSE
    \STATE Regenerate a new $\xi_{m,i}$ assignment plan.
\ENDIF 
\STATE Set the initial temperature $T = 100$, minimum temperature $T_{\text{min}}$, cooling rate $\alpha$, initial iteration number $n = 0$, and maximum iteration number $n_{\text{max}}$. Define the best objective value \eqref{opt2a} as $B_{\text{best}} = 0$, and the optimal vehicle assignment as $\xi_{\text{best}}$.
\REPEAT
    \STATE With the current assignment $\xi_{m,i}^{\text{init}}$, solve the optimization problem \eqref{opt2} and store the result of objective function \eqref{opt2a} as $B_n$.
    \IF{the optimization problem \eqref{opt2} is infeasible}
        \STATE Regenerate a new $\xi_{m,i}$ assignment, and add the current $\xi_{m,i}^{\text{init}}$ to the tabu list $\psi$.
    \ENDIF
    \STATE Generate a neighboring solution of $\xi_{m,i}^{\text{init}}$, denoted by $\xi_{m,i}^{\text{new}}$.
    \STATE Solve the optimization problem \eqref{opt2} with $\xi_{m,i}^{\text{new}}$ and store the result of \eqref{opt2a} as $B_n^{\text{new}}$.
    \vspace{0.1cm}
    \IF{$B_n^{\text{new}} \geq B_n$ \OR $\text{rand} \leq e^{\frac{B_n^{\text{new}} - B_n}{T}}$} 
        \STATE Update $\xi_{m,i}^{\text{init}} = \xi_{m,i}^{\text{new}}$ and $B_n = B_n^{\text{new}}$.
        \IF{$B_n \geq B_{\text{best}}$}
            \STATE Update $\xi_{\text{best}} = \xi_{m,i}^{\text{init}}$ and $B_{\text{best}} = B_n$. 
        \ENDIF
    \ENDIF
    \STATE Update the temperature: $T = \alpha T$ and increment iteration count: $n = n + 1$.
\UNTIL{$n_{\text{max}}$ is reached \OR $T \leq T_{\text{min}}$.}
\end{algorithmic}
\end{algorithm}

The inner optimization problem addresses the optimization of the remaining variables in $\chi$. To solve the inner optimization problem, we propose using the alternating optimization method, and the details are given in \textbf{Algorithm \ref{alg2}}. If the inner optimization problem fails, for example, due to violations of the constraints \eqref{opt1e} or \eqref{opt1f}, the current vehicle assignment plan, $[\xi_{m,1,i}, \xi_{m,2,i}, \ldots, \xi_{m,K,i}]$, becomes infeasible. In such cases, the outer optimization problem must be re-executed. This re-execution involves selecting a new vehicle assignment plan $\xi_{m,k,i}$, while the previous, infeasible assignment plan is added to a tabu list $\psi$ to prevent repeated exploration of the same infeasible solution. This iterative process continues until both the outer and inner optimization problems converge to a feasible solution, satisfying all constraints. By this mechanism, the algorithm dynamically adjusts the vehicle assignments and resource allocations, balancing computational loads across RSUs while ensuring optimal communication and sensing performance. The convergence behavior of \textbf{Algorithm \ref{alg2}} is shown in Appendix~\ref{coverge Appendix}.

\begin{algorithm}
\caption{Iterative Sensing, Communication, and Semantic Optimization Algorithm}\label{alg2}
\begin{algorithmic}[1]
\STATE Set the iteration number $l = 1$, and initialize $\mathbf{A}_{m,k,i}^{\left(0\right)} = \mathbf{I}$ and $\rho_{m,k,i}^{\left(0\right)}$.
\REPEAT
    \STATE Fix $\mathbf{A}_{m,k,i} = \mathbf{A}_{m,k,i}^{\left(l-1\right)}$ and $\rho_{m,k,i} = \rho_{m,k,i}^{\left(l-1\right)}$, and use the optimal vehicle assignment $\xi_{\text{best}}$ obtained from \textbf{Algorithm \ref{alg HH}} to solve \eqref{opt2} for the variables \footnotesize$\left( \mathbf{W}_{m,k,i}^{\left(l\right)}, \mathbf{R}_{m,k,i}^{\left(l\right)}, \eta_{S_{m,k,i}}^{\left(l\right)}, \eta_{m,k,i}^{\left(l\right)}, f_{m,k,i}^{\left(l\right)}, \mathbf{\Omega}_{m,k,i}^{\left(l\right)} \right)$. \normalsize
    \vspace{0.1cm}
    \STATE Fix $\mathbf{W}_{m,k,i} = \mathbf{W}_{m,k,i}^{\left(l\right)}$, $\mathbf{R}_{m,k,i} = \mathbf{R}_{m,k,i}^{\left(l\right)}$, and $\rho_{m,k,i} = \rho_{m,k,i}^{\left(l-1\right)}$, then solve \eqref{opt2} to update $\left( \mathbf{A}_{m,k,i}^{\left(l\right)}, \eta_{S_{m,k,i}}^{\text{new}} \right)$.
    \vspace{0.1cm}
    \STATE Use the bisection method to find the updated semantic extraction ratio $\rho_{m,k,i}^{\left(l\right)}$.
    \vspace{0.1cm}
    \STATE Increment iteration: $l = l + 1$.
    \vspace{0.1cm}
\UNTIL{$\left|\sum_{k=1}^K \eta_{S_{m,k,i}}^{\left(l\right)} - \eta_{S_{m,k,i}}^{\text{new}} \right| < \epsilon$.}
\STATE Apply Gaussian randomization to find rank-one solutions for the beamforming matrices.
\end{algorithmic}
\end{algorithm}

To summarize the overall procedure, including the application of the particle filter and the associated optimization problems, we present the detailed workflow in \textbf{Algorithm \ref{alg summ}}.

\begin{algorithm}[t]
    \caption{A Complete Procedure of ISCSC Design for DT-enabled Vehicular Networks.}
    \label{alg summ}
    \begin{algorithmic}[1]
        \STATE Generate $K$ vehicles and $M$ RSUs in the system.
        \STATE Conduct steps 1-3 in \textbf{Algorithm \ref{alg:pf}}.
        \STATE Apply \eqref{eq55} to predict the state and obtain the measurements.
        \STATE Execute \textbf{Algorithm \ref{alg HH}} and \textbf{Algorithm \ref{alg2}}.
        \STATE Conduct steps 4-11 in \textbf{Algorithm \ref{alg:pf}}.
    \end{algorithmic}
\end{algorithm}

The per-iteration complexity of Algorithm \ref{alg HH} is $\mathcal{O}\left(KN_t \log \left(K N_t\right) \right)$. The per-iteration complexity of Algorithm \ref{alg2} is $\mathcal{O} \left(K^2 N_t^2 \right)$.

\section{Numerical Results}

In this section, we present numerical results to assess the efficacy of the proposed designs. Our setup assumes that antennas are half-wavelength spaced. The values of key parameters used in this paper are listed in Table.~\ref{table 1}. The NF coverage is approximately 300 meters. For simplicity, we consider \eqref{dt_RCRB2} as a multi-variable linear equation. The positions of vehicles follow a Poisson distribution over a highway segment measuring 100 meters in length and 10 meters in width. Each vehicle is modeled with a length of $4.5$ m and a width of $2$ m, and the minimum distance between vehicles is $3.5$ m.

\begin{table}[!t]
    \centering
    \caption{List of Simulation Parameters.}
    \label{table 1}
        \begin{tabular}{|c|c|} 
             \hline
             \textbf{Symbol} & \textbf{Value} \\ 
             \hline 
             \hline
             $N_t$ & 310 \\ 
             \hline
             $N_r$ & 3 \\
             \hline
             $M$ & 2\\
             \hline
             $K$ & 5\\
             \hline
             $\Delta T$ & 0.02 s\\
             \hline
             $P_t$ & 25 dBm \\
             \hline
             $\iota$ & 1.1 \\
             \hline
             $\rho$ & 0.81\\
             \hline
             $F$ & 10 \\
             \hline
             $\mathbf{Q}_1$ \cite{yuan2020bayesian,ding2024joint, liu2022learning} & $ [0.02, 0.3, 1, 0.1] $\\
             \hline
             $\mathbf{Q}_2$ \cite{yuan2020bayesian,ding2024joint, liu2022learning}& $[0.04, 0.06, 1]$ \\
             \hline
             $f$ & $50$ GHz\\
             \hline
             $C_{m,k,i}$ & $[1,2] \times 10 ^3 $ cycles/bit\\
             \hline
             $D_{m,k,i}$ \cite{do2022digital} & $[1,3] \times 10^3$ bits\\
             \hline
             $T^{\text{max}}$ & 0.015 s\\
             \hline
             $F^{\text{max}}$  \cite{gong2022resource}& 5.8 GHz\\
             \hline
             $\sigma^2_c, \sigma^2_r$ & -30 dBm\\
             \hline
             $\epsilon$ & 0.1\\
             \hline
             $\kappa$ \cite{do2022digital, van2022edge, liu2021digital, huang2023mobile} & $10^{-28}$\\
             \hline
             $\varepsilon$ & 0.5 \\
             \hline
             $[\nu_1, \nu_2]$ & $[243.2\;\text{MHz/degree}, 121.6\;\text{MHz/m}]$\\
             \hline
             $\nu_3$ & $\nu_3 \sim \mathcal{CN}\left(0,1000\right)$\\
             \hline
         \end{tabular}
\end{table}

\subsection{Tracking and Digital Twin Performances}

\begin{figure*}[!t]
    \centering
    \subfloat[Angle RMSE]{%
        \includegraphics[width=.40\linewidth]{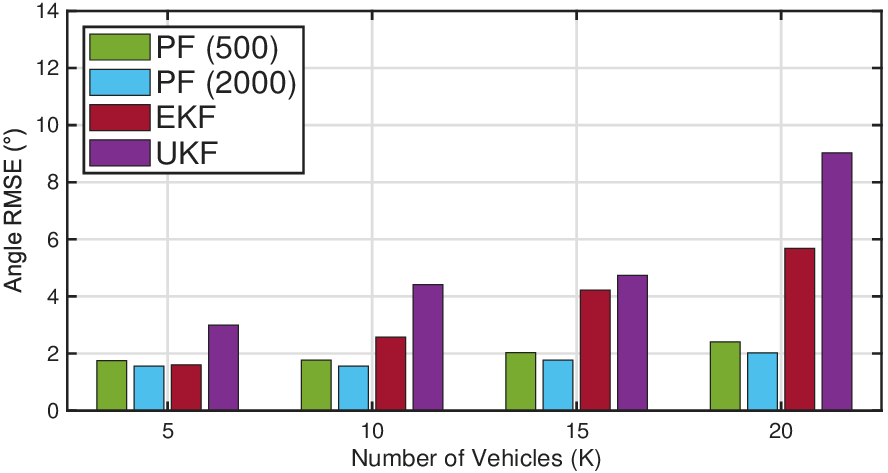}}
    \subfloat[Range RMSE]{%
        \includegraphics[width=.40\linewidth]{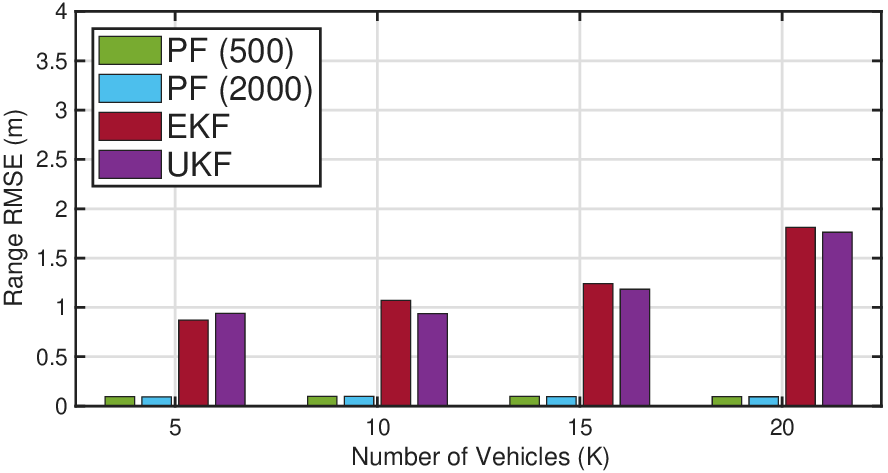}}\\
    \subfloat[Velocity RMSE]{%
        \includegraphics[width=.40\linewidth]{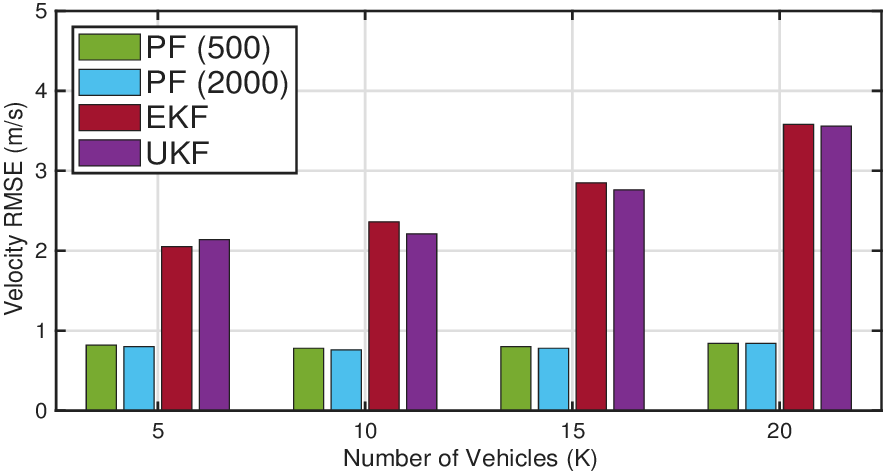}}
    \subfloat[Average computing time]{%
        \includegraphics[width=.40\linewidth]{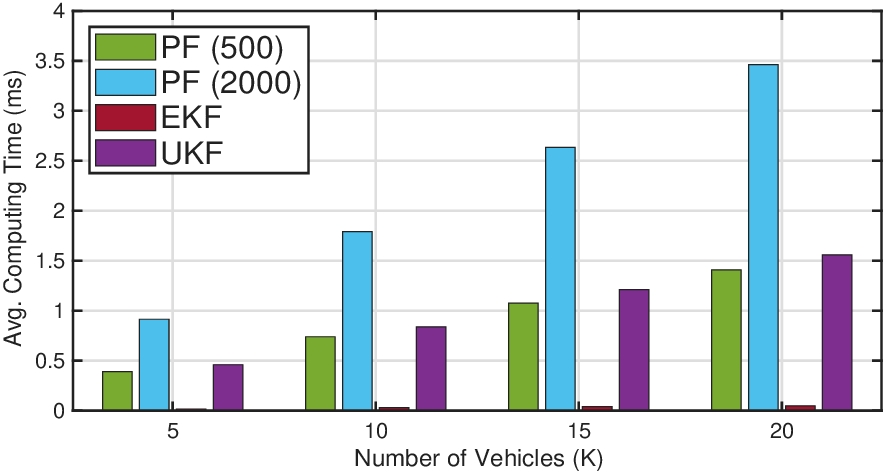}}
    \caption{Tracking performance and Computing time versus the number of vehicles.}
    \label{tracking_performance}
\end{figure*}

To evaluate the tracking performance under different vehicular densities, we compare the PF using $N_s = 500$ and $N_s=2000$ with the EKF \cite{meng2023vehicular} and unscented Kalman filter (UKF) \cite{xu2024isac}. Figs.~\ref{tracking_performance}(a)–(c) report the angle, distance, and velocity root mean square error (RMSE) as the number of vehicles grows, while Fig.~\ref{tracking_performance}(d) shows the corresponding average computation time for each filter.

Fig.~\ref{tracking_performance}(a) shows that the PF achieves the lowest angle RMSE across all vehicle densities. Even with only 500 particles, the PF outperforms both EKF and UKF, and the performance marginally improves with 2000 particles. The EKF and UKF experience rapidly increasing errors as $K$ grows, reflecting their vulnerability to nonlinear models. In contrast, the PF remains stable because it can represent nonlinear and non-Gaussian posterior distributions through its weighted particles.

A similar trend is observed in Fig.~\ref{tracking_performance}(b), which illustrates the range RMSE. The PF consistently attains the highest accuracy across all vehicle densities, maintaining an RMSE below $0.5$~m even at $K=20$. Increasing the number of particles from $500$ to $2000$ yields only a marginal improvement, indicating that using 500 particles is already effective. In contrast, both the EKF and UKF exhibit substantially higher RMSE, with performance degradation becoming more pronounced as the number of vehicles increases. Fig.~\ref{tracking_performance}(c) demonstrates the velocity RMSE, where the PF again provides the most accurate estimates, whereas the EKF and UKF incur markedly larger errors.

Fig.~\ref{tracking_performance}(d) compares the average computational cost. As expected, the PF with 2000 particles incurs a higher computation time than both the EKF and UKF. Nevertheless, the PF with 500 particles achieves a competitive computation time (below 2~ms even when $K=20$) while still delivering substantially better tracking accuracy than the EKF and UKF. This computation time also satisfies the latency requirement discussed later. Although the EKF provides the lowest computational overhead, the PF with 500 particles offers a more favorable accuracy–complexity trade-off within the allowable time budget. This observation justifies our choice of adopting the PF in the proposed framework.

\begin{figure}[!t]
    \centering
    \subfloat[X-coordinate RMSE]{%
        \includegraphics[width=0.85\linewidth]{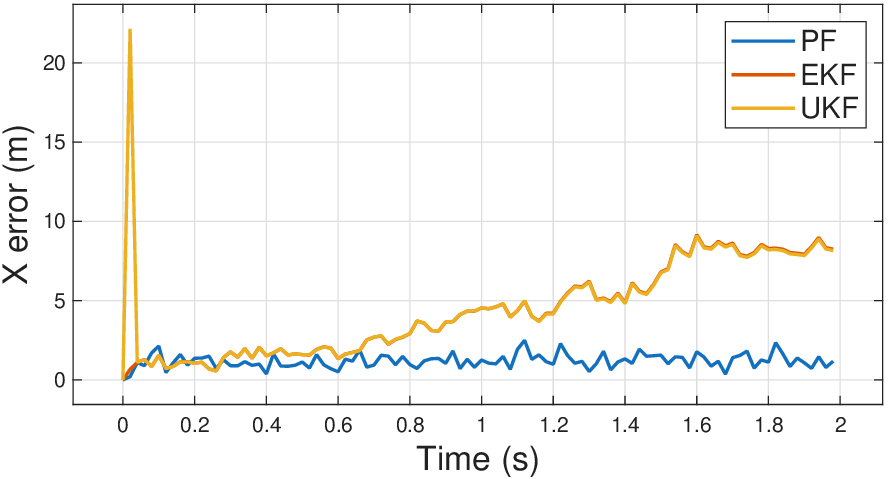}}\\
    \subfloat[Y-coordinate RMSE]{%
        \includegraphics[width=0.85\linewidth]{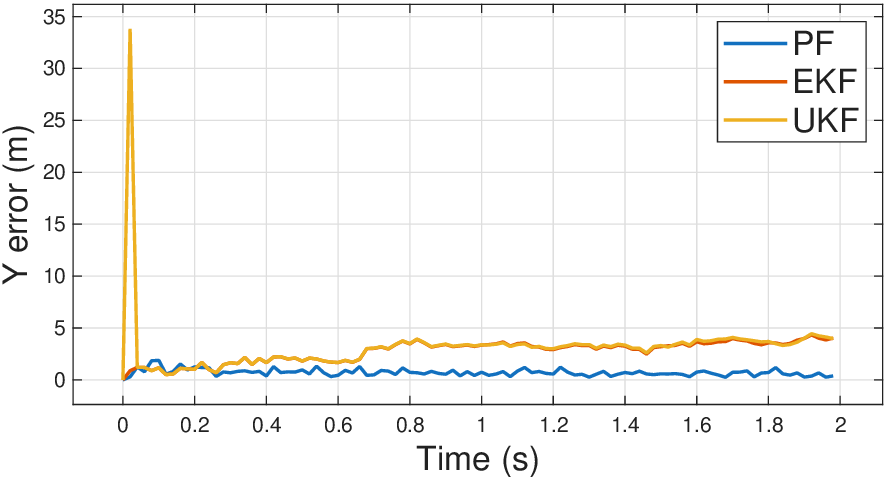}}
    \caption{The X and Y coordinated RMSE for the generated DT model with $K=5$.}
    \label{dt plot}
\end{figure}

Fig.~\ref{dt plot} illustrates the evolution of the average DT modeling error for the X and Y coordinates when 5 vehicles are present. The PF with $N_s = 500$, EKF, and UKF are compared. As shown in Fig.~\ref{dt plot}(a), the PF maintains consistently low X coordinate RMSE throughout the simulation interval, whereas the EKF and UKF yield substantially larger errors, with the UKF exhibiting a pronounced initial spike before gradually stabilizing. A similar trend is observed in Fig.~\ref{dt plot}(b), where the PF sustains lower Y coordinate errors, while the EKF and UKF accumulate increasing deviations over time. Overall, the PF delivers the highest-fidelity DT reconstruction, achieving significantly lower coordinate errors than both EKF and UKF.

\subsection{Semantic Communication and Sensing Performances}

\begin{figure}[!t]
    \centering
    \includegraphics[width=0.85\linewidth]{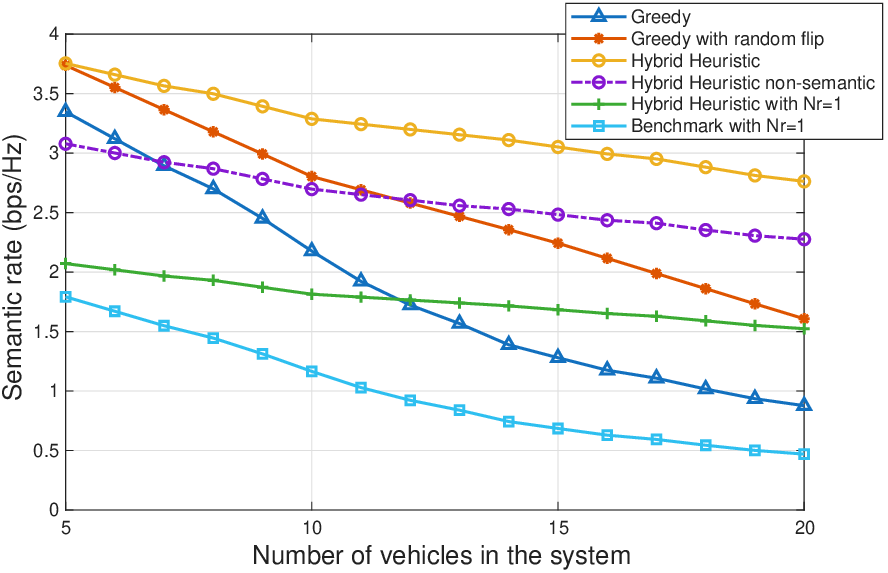}
    \caption{Average semantic transmission rate against the number of vehicles in the system.}
    \label{tr vehicles}
\end{figure}

\begin{figure}[!t]
    \centering
    \includegraphics[width=0.85\linewidth]{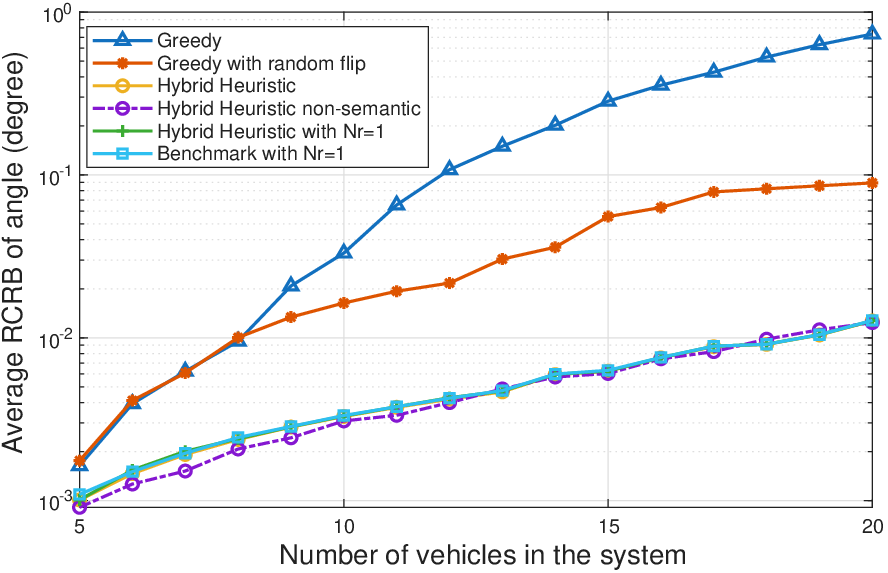}
    \caption{Average RCRB of angle against the number of vehicles in the system.}
    \label{tr angle}
\end{figure}

\begin{figure}[!t]
    \centering
    \includegraphics[width=0.85\linewidth]{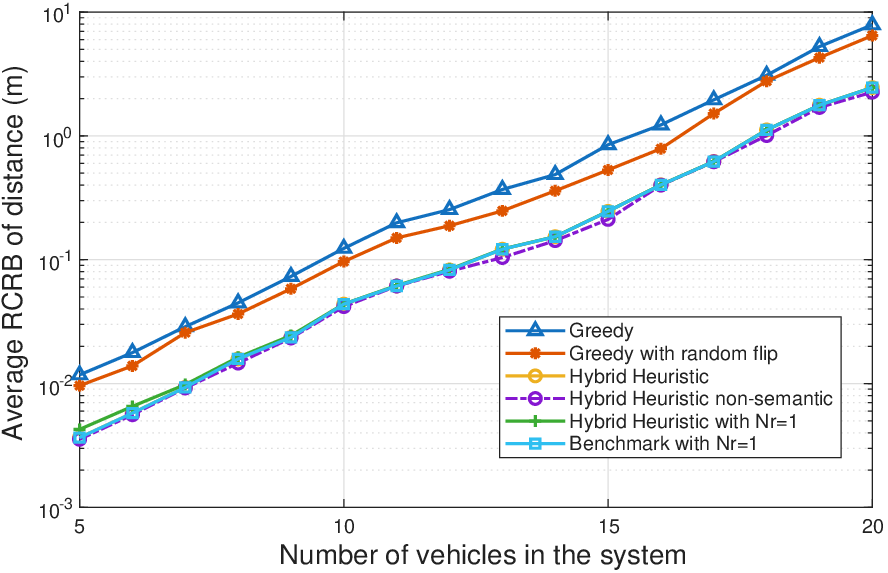}
    \caption{Average RCRB of distance against the number of vehicles in the system.}
    \label{tr distance}
\end{figure}

In Figs. \ref{tr vehicles}, \ref{tr angle}, and \ref{tr distance}, we evaluate the system's performance by varying the number of vehicles. Two vehicle assignment benchmarks are used in the evaluation: the first is the greedy algorithm, and the second is the greedy algorithm with random flip, which randomly flips some values in the solution found by the greedy algorithm. Fig. \ref{tr vehicles} illustrates the average semantic rate versus the number of vehicles in the system. As the number of vehicles increases from 5 to 20, the semantic rate gradually decreases for all schemes due to the limited power budget and increasing multi-vehicle interference, which reduces the resources available per vehicle. The greedy algorithm exhibits the steepest decline, starting at approximately 3.3 bps/Hz with 5 vehicles, and dropping to below 1.0 bps/Hz when the number of vehicles reaches 20. This confirms the greedy method’s vulnerability to sub-optimal vehicle assignments, leading to poor performance. Introducing randomness to the greedy algorithm, labeled as ``greedy with random flip’’ in the figure, slightly improves the performance, with the semantic rate decreasing more gracefully from about 3.7 bps/Hz at 5 vehicles to roughly 1.5 bps/Hz at 20 vehicles, demonstrating improved robustness against sub-optimal solutions. In contrast, the proposed HH algorithm achieves consistently superior performance. It starts at around 3.8 bps/Hz for 5 vehicles and maintains approximately 2.8 bps/Hz even at 20 vehicles, showing the slowest rate degradation compared to the benchmark vehicle assignment algorithms. This robustness highlights HH’s effectiveness in optimizing vehicle assignment. With the HH algorithm, when semantic communication is not used, the average semantic rate drops by 18\%. To further compare the proposed beamforming design with the existing ISAC beamforming designs, we reduce the number of receive antennas to a single antenna, i.e., multi-input-single-output (MISO). This configuration enables a direct comparison of our design with existing ISAC beamforming methods such as \cite{wang2023near}, which is presented in the figure as ``Benchmark with $N_r = 1$''. When the number of vehicles in the system is small, the proposed algorithm achieves a semantic rate that is similar to the benchmark, indicating that both methods perform effectively under lightly loaded conditions. However, as the vehicle number increases, the benchmark curve exhibits a noticeable decline, whereas the proposed design decreases more gradually, demonstrating improved robustness to congestion. Moreover, unlike the benchmark method, which is limited to MISO configurations, the proposed beamforming framework seamlessly accommodates both MISO and MIMO settings, providing greater flexibility and scalability for practical vehicular networks.

Fig.~\ref{tr angle} illustrates the average RCRB for angle estimation as the number of vehicles increases. As shown, the greedy algorithm exhibits the weakest performance. Although it achieves an RCRB on the order of $10^{-2}$ degrees when only 5 vehicles are present, its accuracy deteriorates rapidly under higher traffic density, reaching approximately $10^{1}$ degrees at 20 vehicles. This sharp increase reflects the poor vehicle assignment capability of the greedy strategy in multi-vehicle scenarios. Incorporating random flips leads to consistently lower RCRB values than the pure greedy scheme. However, its performance still degrades noticeably as the number of vehicles grows. In contrast, the proposed HH algorithm achieves markedly superior angular sensing accuracy, with its RCRB increasing only mildly from roughly $10^{-3}$ degrees at 5 vehicles to about $10^{-2}$ degrees at 20 vehicles. The non-semantic HH design exhibits performance trends consistent with the semantic-enabled HH design, confirming that incorporating semantic information does not compromise angular sensing accuracy. Finally, the proposed beamforming design with $N_r=1$ delivers performance comparable to the benchmark method, while simultaneously offering improved semantic rate, as evidenced in Fig.~\ref{tr vehicles}.

Fig.~\ref{tr distance} presents the average RCRB of distance estimation as the number of vehicles increases. The greedy algorithm again performs the worst, with its RCRB escalating sharply from approximately $10^{-2}$ m at 5 vehicles to nearly $10$ m at 20 vehicles. Introducing random flips yields moderate improvements but still results in a pronounced rise in RCRB. In contrast, the proposed HH algorithm achieves the highest accuracy, with its RCRB increasing only gradually from roughly $5\times10^{-3}$ m at 5 vehicles to below $0.5$ m at 20 vehicles. The non-semantic HH design performs similarly to the semantic-based design, indicating that incorporating semantic information does not distort distance sensing accuracy. Finally, the proposed beamforming design with $N_r = 1$ and the benchmark MISO design exhibit comparable behavior. However, unlike the MISO benchmark, the proposed beamforming design achieves a higher semantic rate and retains the flexibility to operate in either MISO or MIMO configurations.

\subsection{Computing Performance}

\begin{figure}[!t]
    \centering
    \includegraphics[width=0.85\linewidth]{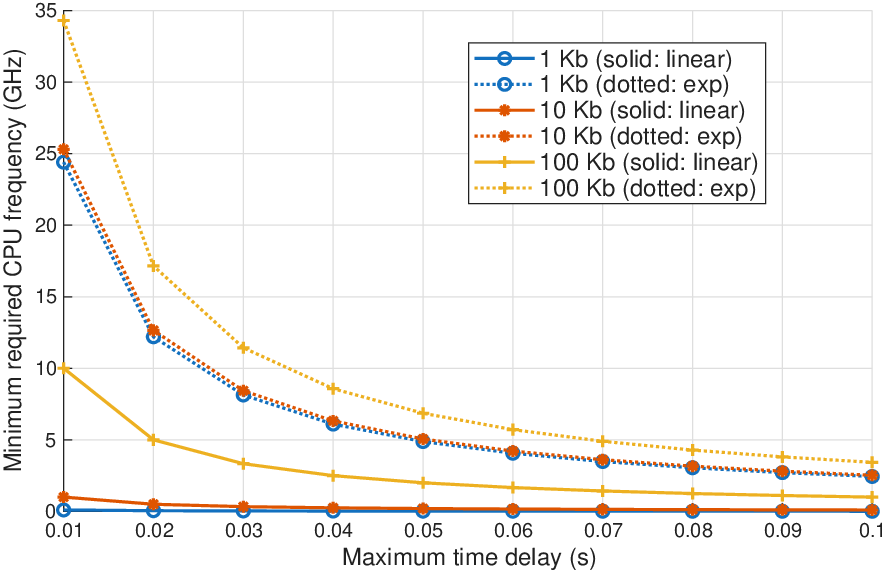}
    \caption{Minimum required CPU frequency versus maximum time delay.}
    \label{time freq}
\end{figure}

\begin{figure}[!t]
    \centering
    \includegraphics[width=0.85\linewidth]{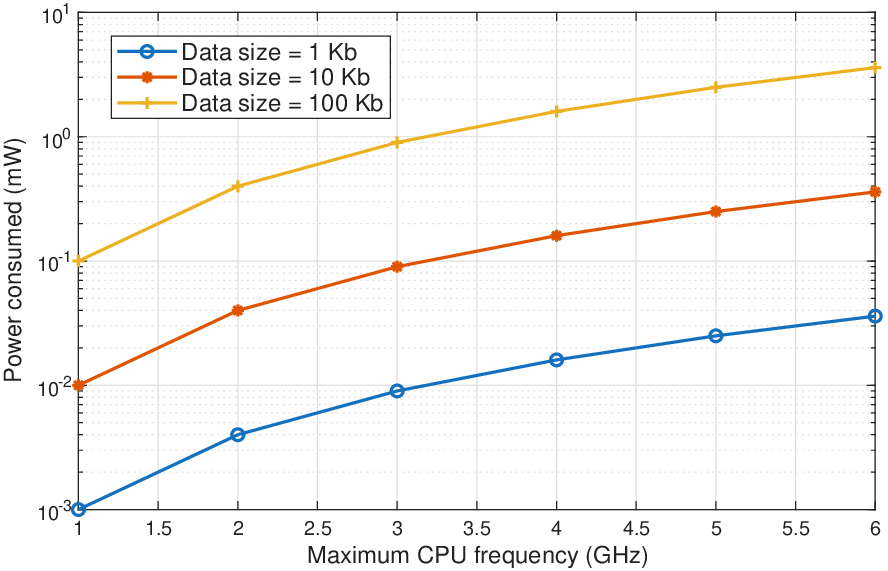}
    \caption{Power consumed versus maximum CPU frequency.}
    \label{power cpu}
\end{figure}

Fig.~\ref{time freq} illustrates the minimum required CPU frequency as a function of the maximum allowable processing time delay for three data sizes. As expected, increasing the allowable time delay significantly reduces the required CPU frequency, since more relaxed latency constraints lower the computation resources needed to process each data block. For any fixed delay, larger data sizes require higher CPU frequencies due to the greater computational workload associated with processing more bits. The figure also compares the linear computational model with an exponential computational model for \eqref{dt_RCRB2}. Under the exponential model, the minimum required CPU frequency grows sharply, especially for strict delay requirements, and can easily exceed the capability of a single processor. In such cases, the system must either increase the maximum tolerable delay or rely on multi-core processing to meet the required computation rate. This highlights the importance of explicitly accounting for sensing error when constructing computational models in latency-sensitive vehicular networks.

Fig. \ref{power cpu} illustrates the relationship between power consumption and maximum CPU frequency for three different data sizes. It is evident that the power consumption rises as CPU frequency increases for all data sizes.

\section{Conclusion and Future Direction}

In this paper, we have proposed an ISCSC framework for DT-enabled vehicular networks, addressing the unexplored integration of ISAC, semantic communication, and near-field effects. In a multi-RSU and MU-MIMO setting, each RSU employs particle filtering for vehicle tracking. We have proposed a hybrid heuristic algorithm that optimally assigns vehicles to RSUs for communication. With the optimal vehicle assignment, we proposed an alternating optimization algorithm to jointly optimize the beamforming matrix, the semantic extraction ratio, and the CPU frequency, ensuring efficient resource allocation and enhancing the semantic transmission rate and sensing accuracy. In the simulation results, we have used performance measures like semantic rate and CRB to confirm that the proposed ISCSC framework outperforms existing designs in terms of semantic throughput and sensing precision, making it a promising approach for future vehicular networks.

Several promising research directions remain open. First, given the limited computational capability at individual RSUs, MEC can be incorporated to offload intensive signal processing and support real-time DT construction. Second, cooperative multi-RSU sensing and distributed information fusion should be explored to enhance sensing accuracy in complicated vehicular environments. Third, extending semantic communication to multi-modal data (e.g., radar and cameras) may yield richer and more robust DT representations. Fourth, the modeling of DT computing latency, including its coefficients and dependence on sensing and tracking errors, should be calibrated and validated using real-world measurements. Fifth, quantifying the performance degradation that arises when far-field models are (incorrectly) applied in inherent near-field regimes represents an important research direction. 
Sixth, advanced signal processing and multi-target tracking techniques for reliably distinguishing and classifying closely spaced vehicles should be developed.
Finally, integrating full-duplex semantic communication with considerations of overhead, interference, and sensing performance constitutes an important direction toward practical large-scale vehicular deployments.

\appendices
\section{Convergence Analysis}
\label{coverge Appendix}
\setcounter{equation}{0}
\numberwithin{equation}{section}

In this appendix, we establish the convergence of the proposed alternating optimization algorithm in Algorithm~\ref{alg2}. By denoting the objective function of the alternating optimization as $O(\cdot)$, and the objective functions in sub-problems by $O_1(\cdot)$, $O_2(\cdot)$ and $O_3(\cdot)$ respectively, we have the following equation:
\begin{equation}
\begin{aligned}
    &\min \; O\left(\mathbf{W}_{m,k,i}, \mathbf{R}_{m,k,i}, \mathbf{A}_{m,k,i}, \rho_{m,k,i} \right) \\
    & = \min \; O_1\left(\mathbf{W}_{m,k,i}, \mathbf{R}_{m,k,i}\right) + \min \; O_2\left(\mathbf{A}_{m,k,i}\right) \\
    & \qquad + \min \; O_3\left(\rho_{m,k,i} \right).
\end{aligned}
\end{equation}

For fixed $\mathbf{A}_{m,k,i}$ and $\rho_{m,k,i}$, the sub-problem with respect to $(\mathbf{W}_{m,k,i},\mathbf{R}_{m,k,i})$ is convex. Letting $l$ denote the iteration index, the optimality of
$(\mathbf{W}_{m,k,i}^{(l)},\mathbf{R}_{m,k,i}^{(l)})$ ensures
\begin{equation}
\begin{aligned}
O_1(\mathbf{W}_{m,k,i}^{(l-1)}, \mathbf{R}_{m,k,i}^{(l-1)})
\ge 
O_1(\mathbf{W}_{m,k,i}^{(l)}, \mathbf{R}_{m,k,i}^{(l)}),
\end{aligned}
\end{equation}
which shows that $O_1(\cdot)$ is monotonically non-increasing. Since \eqref{opt2} is bounded and convex, $O_1(\cdot)$ converges.

For fixed $(\mathbf{W}_{m,k,i},\mathbf{R}_{m,k,i},\rho_{m,k,i})$, the variable $\mathbf{A}_{m,k,i}$ is updated, and it optimal solution $\mathbf{A}_{m,k,i}^{(l)}$ satisfies
\begin{equation}
\begin{aligned}
O_2(\mathbf{A}_{m,k,i}^{(l-1)})
\ge 
O_2(\mathbf{A}_{m,k,i}^{(l)}),
\end{aligned}
\end{equation}
implying that each $\mathbf{A}_{m,k,i}$ update decreases or maintain the objective value. Hence, $O_2(\cdot)$ converges.

For fixed $(\mathbf{W}_{m,k,i},\mathbf{R}_{m,k,i},\mathbf{A}_{m,k,i})$, the scalar $\rho_{m,k,i}$ is obtained using a standard bisection method. Since bisection monotonically refines the feasible interval, the corresponding objective values satisfy
\begin{equation}
\begin{aligned}
O_3(\rho_{m,k,i}^{(l-1)})
\ge
O_3(\rho_{m,k,i}^{(l)}),
\end{aligned}
\end{equation}
thus $O_3(\cdot)$ also converges.

Combining the above results, we have
\begin{equation}
    O^{(l)} = O(\mathbf{W}_{m,k,i}^{(l)},\mathbf{R}_{m,k,i}^{(l)},
\mathbf{A}_{m,k,i}^{(l)},\rho_{m,k,i}^{(l)}) \leq O^{(l-1)},
\end{equation}
is monotonically non-increasing. Since problem~\eqref{opt2} is bounded, the AO algorithm converges to a stationary point.

\bibliographystyle{ieeetr}
\bibliography{ref}

\end{document}